# Marco IA593: Modelo de Gobernanza, Ética y Estrategia para la Integración de la Inteligencia Artificial en la Educación Superior del Ecuador

Luis Chamba-Eras[1], Oscar Miguel Cumbicus Pineda[1], Edison Leonardo Coronel Romero[1], Jessica Katherine Gaona Alvarado[2], Luis Rodrigo Barba Guamán[3]

[1]Universidad Nacional de Loja, Facultad de la Energía, las Industrias y los Recursos Naturales No Renovables, Carrera de Computación, GITIC, Av. Pío Jaramillo Alvarado y Av. Reinaldo Espinosa, código postal 110111, Loja, Ecuador.
[2]Universidad Nacional de Loja, Facultad Agropecuaria y de Recursos Naturales Renovables, Carrera de Ingeniería Ambiental, Av. Pío Jaramillo Alvarado y Av. Reinaldo Espinosa, código postal 110111, Loja, Ecuador.
[3]Universidad Técnica Particular de Loja, Facultad de Ingenierías y Arquitectura, Ciencias de la Computación y Electrónica, GI2A2, San Cayetano Alto, código postal 1101608, Loja, Ecuador.

[lachamba,oscar.cumbicus,edisoncor, jessica.gaona]@unl.edu.ec
lrbarba@utpl.edu.ec

## Resumen

La integración de la Inteligencia Artificial (IA) en las Instituciones de Educación Superior (IES) de Ecuador no es una opción tecnológica, sino un imperativo estratégico para evitar la obsolescencia institucional y la irrelevancia académica en América Latina. Este trabajo presenta el Marco de Trabajo IA593, un modelo de gobernanza, ética y operatividad diseñado para la Universidad Nacional de Loja (UNL), pero escalable como referente para el sistema de educación superior ecuatoriano. El contexto actual revela una urgencia crítica: el Índice Latinoamericano de Inteligencia Artificial 2025 clasifica a Ecuador como un "adoptante tardío en despertar", evidenciando brechas estructurales severas donde la inversión en I+D es apenas del 0.44% del PIB y la producción científica en IA representa una fracción marginal del total global. Aunque existe una Estrategia Nacional para el Fomento de la IA que exige una gobernanza multisectorial, las universidades enfrentan un vacío de normativas internas que regulen el uso de la IA Generativa, poniendo en riesgo la integridad académica y la privacidad de datos. El Marco IA593 responde a este desafío mediante una estructura de cinco pilares interconectados, alineados con los principios FATE (Equidad, Rendición de Cuentas, Transparencia y Ética) y la Recomendación sobre la Ética de la IA de la UNESCO: 1) Gobernanza Transversal (STEER): Establecimiento del Comité de Ética de IA (CEIA) y protocolos de gestión de riesgos para cumplir con la normativa nacional; 2) Docencia y Formación (LEARN): Implementación de la Alfabetización en IA (AI Literacy) como competencia obligatoria y rediseño de la evaluación para mitigar el plagio automatizado; 3) Investigación (DISCOVER): Fomento de la AI for Research (AI4R) para potenciar la producción científica y regular la declaración de uso de herramientas en artículos científicos/académicos; 4) Vinculación (CONNECT): Aplicación de la IA para resolver problemas territoriales y avanzar en los Objetivos de Desarrollo Sostenible; y 5) Gestión (OPERATE): Modernización de la infraestructura tecnológica sostenible y automatización de procesos administrativos. La implementación de este marco permitirá a las IES transitar de un consumo pasivo de tecnología a una adopción soberana y crítica, garantizando el cumplimiento del Reglamento de Régimen Académico del Consejo de Educación Superior y posicionando a la UNL como un actor clave en la reducción de la brecha digital y la fuga de talentos que afecta a la región, y de manera particular a Ecuador.

## 1. Introducción

La irrupción de la Inteligencia Artificial (IA), y específicamente de la IA Generativa (IAGen), ha marcado un punto de inflexión en la historia de la educación superior, comparable a revoluciones tecnológicas precedentes como la electricidad o la internet (A. Rivas et al., 2023). Esta tecnología de propósito general no solo redefine los métodos de enseñanza y aprendizaje mediante la personalización y la tutoría inteligente, sino que altera fundamentalmente la producción científica, la gestión administrativa y la vinculación con la sociedad (Gobierno de Chile, 2021; Universidad Europea, 2024). Sin embargo, la velocidad de adopción de estas herramientas por parte del alumnado ha superado la capacidad de respuesta normativa de las instituciones, generando un escenario de incertidumbre donde conviven oportunidades de innovación con riesgos éticos críticos (Díaz-Noguera et al., 2025; Serrano Acitores, 2025).

A nivel global, la UNESCO y la OCDE han instado a los Estados a establecer marcos de gobernanza que aseguren una IA centrada en el ser humano, transparente y equitativa. No obstante, en América Latina y el Caribe, la adopción se enfrenta a brechas estructurales significativas. Según el Índice Latinoamericano de Inteligencia Artificial 2025 (ILIA 2025), la región contribuye con apenas el 0,21% de las patentes globales de IA y enfrenta déficits severos en infraestructura de cómputo y talento humano especializado (CENIA, 2025; Molina & Medina, 2025). Si bien países como Chile y Brasil han avanzado en estrategias nacionales robustas, Ecuador se encuentra en una etapa de transición, clasificado como un “adoptante” que requiere fortalecer su gobernanza y sus capacidades de investigación para no quedar rezagado en la economía del conocimiento (MINTEL, 2026; UNESCO, 2025a).

En el contexto ecuatoriano, la ausencia de una regulación universitaria específica ha dejado a docentes y gestores en un vacío operativo. Aunque la Estrategia para el Fomento del Desarrollo y Uso Ético y Responsable de la Inteligencia Artificial en el Ecuador (EFIA-EC) establece la necesidad de una gobernanza multisectorial, las Instituciones de Educación Superior (IES) carecen de instrumentos propios que aterricen estos mandatos en el aula y el laboratorio (MINTEL, 2026). El uso no regulado de la IA plantea amenazas directas a la integridad académica —como el plagio automatizado y la falta de verificación de fuentes— y riesgos sobre la privacidad de los datos estudiantiles, la “ratificación” de la educación y la perpetuación de sesgos algorítmicos provenientes del Norte Global (Fábrega Lacoa et al., 2024; Giandana Gigena et al., 2024; UNESCO, 2023a).

Ante este escenario, este trabajo presenta el Marco de Trabajo IA593 (Figura 1), diseñado como una respuesta institucional sistémica para la Universidad Nacional de Loja (UNL) y como modelo orientativo para el sistema de educación superior del Ecuador. A diferencia de enfoques fragmentados que se limitan a prohibir o permitir herramientas, el IA593 propone una estructura de gobernanza integral basada en cinco pilares: Gobernanza (STEER), Docencia (LEARN), Investigación (DISCOVER), Vinculación (CONNECT) y Gestión (OPERATE).

El objetivo de este trabajo es doble: primero, fundamentar teórica y legalmente la necesidad de este marco, alineándolo con la normativa del Consejo de Educación Superior (CES) y los estándares éticos de la UNESCO; y segundo, proporcionar una hoja de ruta operativa para que las universidades transiten de un consumo pasivo de tecnología a una adopción soberana, ética y estratégica, capaz de potenciar la investigación (AI4R) y reducir las brechas digitales en el territorio (PNUD, 2025a; UNESCO, 2023b).

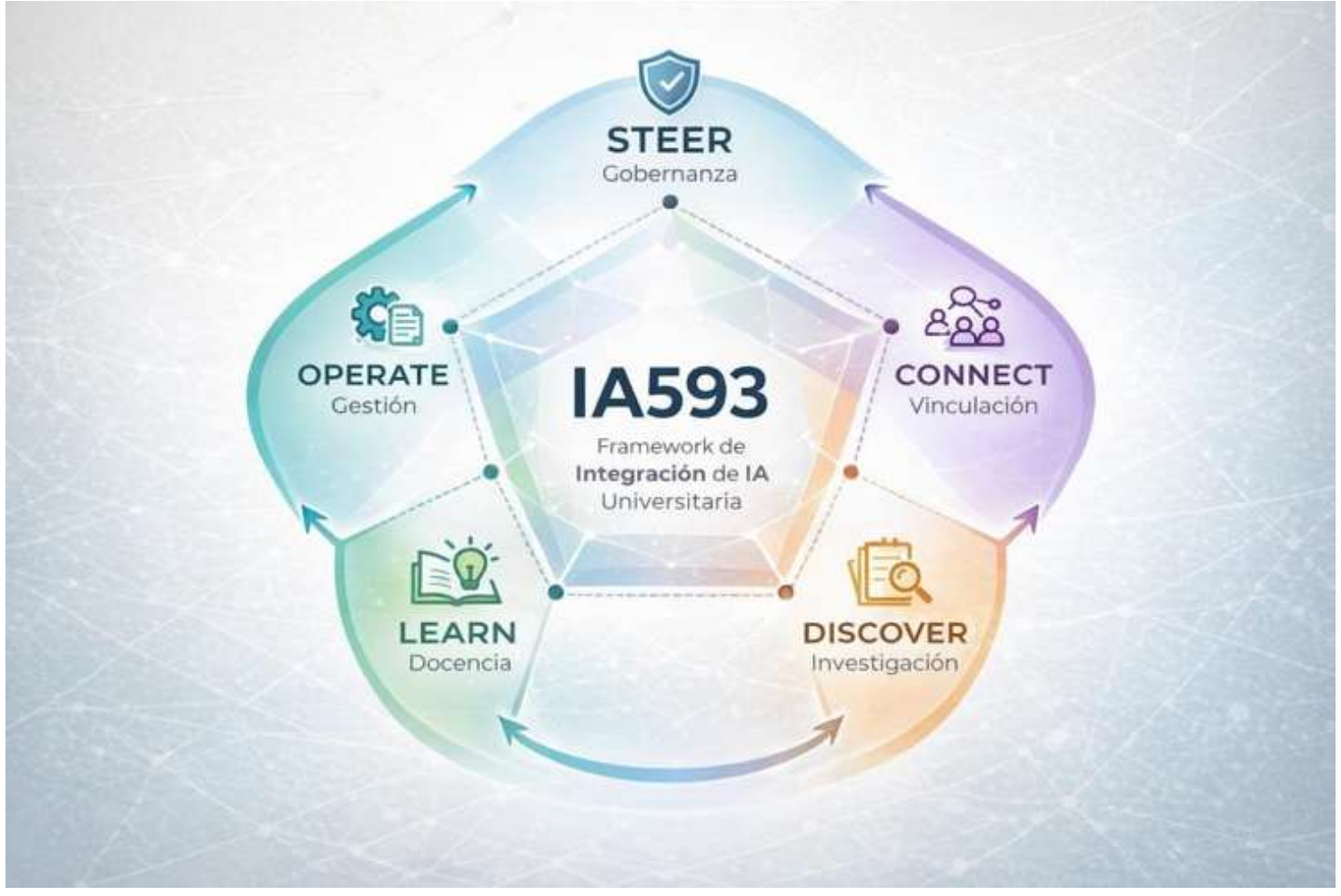

**Figura 1.** Marco de Trabajo IA593 (generado con Gemini).

## 2. Antecedentes y marco contextual

La formulación del Marco de Trabajo IA593 no surge de la arbitrariedad, sino de la imperiosa necesidad de alinear la gestión universitaria con la realidad geopolítica, la estrategia nacional de desarrollo tecnológico y la normativa vigente de educación superior en Ecuador. A continuación, se detalla el ecosistema actual y los habilitantes legales que justifican este instrumento.

### 2.1. Diagnóstico de madurez

El contexto nacional ha sido diagnosticado con precisión por la Evaluación del Estadio de Preparación (RAM) realizada por la UNESCO en 2025. Este informe revela que Ecuador posee un marco de gobernanza digital en desarrollo, pero carece de una estrategia de IA consolidada y financiada. Se identifican brechas críticas en la "Competencia 28" del Currículo Nacional, la cual no integra suficientemente el pensamiento computacional ni la ética de la IA, dejando a la educación superior con la responsabilidad de nivelar estas competencias (UNESCO, 2025a).

Asimismo, el informe destaca que, aunque Ecuador cuenta con una Ley de Protección de Datos Personales (LOPDP) desde el 2021 (MINTEL, 2021) y un Reglamento General de la Ley Orgánica de Protección de Datos Personales (RGLOPDP) desde el 2023 (MINTEL, 2023), existe un vacío en normativas secundarias para la adquisición pública de sistemas de IA, lo que expone a las universidades públicas a riesgos al contratar software educativo sin auditoría de sesgos o privacidad (MINTEL, 2026; UNESCO, 2025a).

### 2.2. La brecha global de innovación

A nivel global, el Informe sobre Tecnología e Innovación 2025 advierte que la IA está exacerbando las desigualdades. La concentración de patentes y modelos fundacionales en el Norte Global obliga a los países en desarrollo a adoptar estrategias de "IA Inclusiva" para evitar una nueva dependencia colonial tecnológica. El Marco IA593 responde a este llamado alineándose con los Objetivos de Desarrollo Sostenible (ODS) y la soberanía digital (Naciones Unidas, 2025).

### 2.3. Ecuador en el ILIA 2025

La justificación del Marco IA593 se ancla en la realidad geopolítica revelada por el ILIA 2025. Este estudio clasifica a Ecuador en la categoría de "adoptante" junto a República Dominicana y Guatemala. Si bien el país ha mostrado una mejora acelerada en factores habilitantes, enfrenta desafíos estructurales que la universidad debe mitigar (CENIA, 2025).

Los indicadores críticos del ILIA 2025 que sustentan la intervención universitaria son:

- Brecha de talento humano especializado: La penetración de habilidades de IA en la fuerza laboral regional es apenas del 0.1%, tres veces inferior al promedio mundial (0.32%). Además, existe una fuga de talento ("brain drain") significativa hacia el Norte Global debido a la falta de ecosistemas locales robustos (CENIA, 2025).
- Déficit de infraestructura de cómputo: La capacidad de procesamiento en la región es desigual. Ecuador posee una capacidad de cómputo limitada en comparación con líderes regionales como Brasil o Chile, lo que restringe la capacidad de entrenar modelos soberanos y aumenta la dependencia de APIs comerciales extranjeras (CENIA, 2025; PNUD, 2025b).
- Concentración de la investigación: El Informe sobre Tecnología e Innovación 2025 alerta que la investigación en IA está hiperconcentrada, EE. UU. y China poseen el 90% de las patentes y publicaciones de alto impacto, dejando a América Latina con menos del 2% de la producción global. Sin embargo, Ecuador destaca positivamente con una participación femenina en investigación de IA del 27%, superior al promedio regional (23.6%), un activo que el Marco IA593 busca potenciar (CENIA, 2025; Naciones Unidas, 2025).

### 2.4. El ecosistema de IA en Ecuador: Diagnóstico y brechas

El posicionamiento de Ecuador en el panorama tecnológico regional es el punto de partida para cualquier estrategia institucional. Según ILIA 2025, Ecuador se clasifica en la categoría de "adoptante" (países con puntajes entre 35 y 60 sobre 100), mostrando un despertar tardío pero acelerado en la región (CENIA, 2025).

Si bien el país ha mostrado avances en la alfabetización digital temprana y en la paridad de género en investigación (con una participación femenina del 27%, superior al promedio regional), el ecosistema enfrenta brechas estructurales críticas que el Marco IA593 busca mitigar:

- Déficit en investigación y desarrollo (I+D): La inversión en I+D es apenas del 0.44% del PIB, una cifra insuficiente que limita la capacidad de innovación endógena y perpetúa la dependencia de tecnologías extranjeras (CENIA, 2025; MINTEL, 2026).
- Baja producción de patentes: La generación de propiedad intelectual vinculada a la IA es marginal, situando al país lejos de los líderes regionales como Brasil o Chile, lo que evidencia una desconexión entre la academia y el sector productivo (CENIA, 2025).
- Infraestructura de cómputo limitada: La capacidad de procesamiento (GPU) per cápita es inferior al promedio regional (26.0 Teraflops frente a 34.2), lo que restringe el desarrollo de modelos de IA complejos y soberanos (MINTEL, 2026).

Este diagnóstico confirma que las IES no pueden permanecer pasivas; deben actuar como catalizadores para cerrar estas brechas mediante marcos de trabajo que fomenten la investigación aplicada (AI4R) y la formación de talento especializado.

## 2.5. Marco legal habilitante

La adopción de la IA en la universidad debe enmarcarse en las políticas de Estado y las leyes de protección de derechos fundamentales.

- Estrategia para el Fomento del Desarrollo y Uso Ético y Responsable de la Inteligencia Artificial en el Ecuador (EFIA-EC): Publicada en el Registro Oficial Suplemento Nº 206 (enero 2026), establece la hoja de ruta nacional. Esta normativa asigna a la Academia un rol protagónico dentro del modelo de gobernanza multiactor, encargándola de "liderar la I+D+i, formar profesionales y desarrollar soluciones inclusivas" (MINTEL, 2026). El Marco IA593 responde directamente a los objetivos de la estrategia nacional, específicamente al Eje 1 (Gobernanza) y Eje 2 (Desarrollo de Talento), operacionalizando el mandato de crear "lineamientos éticos, transparentes y responsables" para el uso de la IA en el territorio nacional.
- Ley Orgánica de Protección de Datos Personales (LOPDP): El uso de plataformas de IA en la educación implica el procesamiento masivo de datos de estudiantes, lo que activa las obligaciones de la LOPDP. El artículo 42 de esta ley exige realizar evaluaciones de impacto cuando el tratamiento de datos pueda generar un alto riesgo para los derechos y libertades, como es el caso del perfilamiento automatizado de estudiantes o el uso de datos biométricos (UNESCO, 2025a). El Marco IA593, a través de su Pilar de Gobernanza, establece los protocolos necesarios para cumplir con este requisito legal, garantizando la privacidad y la autodeterminación informativa de la comunidad universitaria.

## 2.6. Normativa de educación superior (CES)

Aunque el CES no cuenta con un reglamento específico para IA, el Reglamento de Régimen Académico (RRA) vigente establece principios que son directamente impactados por estas tecnologías y que el Marco IA593 busca salvaguardar:

- Definición de investigación (Art. 3 y 34): El RRA define la investigación como una labor "creativa, sistemática y sistémica" desarrollada bajo "principios éticos". El uso no regulado de IA Generativa para la redacción automática de textos o la fabricación de datos vulnera la naturaleza creativa y ética exigida por la norma (CES, 2023).
- Integridad y fraude académico (Art. 38): La normativa tipifica el fraude académico como la "apropiación de ideas" o el uso de soportes no autorizados. Sin una regulación clara como la que propone el IA593, el uso de herramientas como ChatGPT en evaluaciones podría constituir una violación masiva al Art. 38, poniendo en riesgo la validez de los títulos otorgados (CES, 2023; Villegas Dianta et al., 2025).
- Recursos de aprendizaje (Art. 61): El CES exige que las IES garanticen recursos tecnológicos adecuados y mecanismos de control de honestidad académica en modalidades en línea (CES, 2023). El IA593 actualiza estos mecanismos para la era algorítmica, asegurando que la tecnología sea un soporte pedagógico y no un atajo al aprendizaje.

## 3. Estado del arte: Tendencias y modelos de referencia

Para garantizar la pertinencia y sostenibilidad del Marco IA593, su diseño se fundamenta en un análisis exhaustivo de los estándares internacionales de gobernanza, las mejores prácticas de universidades referentes en Iberoamérica, trabajos relacionados y la evidencia cuantitativa sobre la brecha científica en la región. A continuación, se exponen las dimensiones que validan la estructura propuesta.

### 3.1. Estándares internacionales: La ética y el control humano

La gobernanza de la IA en la educación superior ha transitado de una etapa de exploración técnica a una de regulación ética normativa. El referente indiscutible es la Recomendación sobre la Ética de la IA de la UNESCO, adoptada por 193 estados miembros. Este instrumento establece que el despliegue de la IA debe estar subordinado a los derechos humanos y, crucialmente, debe garantizar el principio de "Human-in-the-loop" (HITL) o supervisión humana significativa (Bellas et al., 2025; UNESCO, 2024).

El enfoque HITL, respaldado también por la OCDE y la Comisión Europea, dicta que la IA no debe usurpar la toma de decisiones críticas en la educación (como la evaluación sumativa o la admisión) sin la intervención y validación final de un docente (Ministry of National Education, 2025). El Marco IA593 operacionaliza este estándar en su Pilar II (Docencia), rechazando el solucionismo tecnológico automático y promoviendo una "interacción humano-IA" donde la tecnología aumenta las capacidades cognitivas sin reemplazar el juicio pedagógico (Bellas et al., 2025; OECD, 2023a).

Asimismo, la tendencia global hacia una "IA Digna de Confianza" (Trustworthy AI) exige transparencia y explicabilidad (XAI). Los modelos educativos no pueden ser "cajas negras"; estudiantes y profesores deben comprender cómo y por qué un sistema de IA genera una recomendación o resultado, un requisito que el IA593 integra en sus protocolos de gobernanza (Bellas et al., 2025).

### 3.2. Benchmarking universitario: La respuesta regional

Ante el vacío regulatorio nacional en muchos países de América Latina, diversas universidades han tomado la iniciativa de crear sus propias normativas. El análisis comparado valida los componentes del IA593 al mostrar coincidencias estratégicas con instituciones líderes:

- Universidad de Las Américas (UDLA - Chile) y la Universidad Andrés Bello (Chile): Han implementado, un "Marco para el uso de la IA" institucionalizado que define lineamientos explícitos para Docencia, Investigación y Vinculación con el Medio (Villegas Dianta et al., 2025); y unos lineamientos para el uso responsable de la IA (Universidad Andrés Bello, 2024). Al igual que el IA593, la UDLA no prohíbe la IA, sino que regula su uso mediante principios de integridad académica, exigiendo la declaración de uso y promoviendo la alfabetización digital obligatoria para docentes y estudiantes.
- Universidad Nacional Autónoma de México (UNAM): A través de grupos de trabajo colegiados, ha emitido recomendaciones que priorizan la ética y la protección de datos personales frente a la IAGen. Su enfoque destaca la necesidad de actualizar los reglamentos de evaluación para evitar el plagio automatizado, alineándose directamente con el componente de "Integridad Académica y Evaluación Crítica" del Marco IA593 (Castañeda de León et al., 2025; Grupo de trabajo de Inteligencia Artificial Generativa de la UNAM, 2025).

- Casos Europeos (Bolonia y Murcia): Universidades como la de Bolonia (Italia) y Murcia (España) han avanzado hacia directrices basadas en riesgos y el uso de Learning Analytics para la personalización, pero siempre bajo un marco de "honestidad académica" y no de vigilancia punitiva, reforzando la visión humanista que propone el IA593 (Serrano Acitores, 2025; WOOCLAP, 2025).

Este benchmarking confirma que el IA593 sigue la tendencia internacional de institucionalizar la gobernanza, moviéndose de la reacción aislada a la estrategia corporativa integral.

### 3.3. La brecha científica y el imperativo AI4R (AI for Research)

La justificación más crítica para el Pilar III (Investigación) del Marco IA593 proviene de la alarmante brecha de productividad científica y tecnológica que separa a América Latina del Norte Global.

- Producción y patentes: Según ILIA 2025, la región representa una fracción marginal de la producción mundial de conocimiento en IA. Mientras China y EE. UU. dominan las publicaciones y patentes, América Latina contribuye con menos del 2% de las publicaciones globales y apenas el 0,21% de las patentes de IA (CENIA, 2025; Molina & Medina, 2025).
- Concentración y fuga de talento humano: La investigación en la región está altamente concentrada en Brasil, México y Chile, dejando a países como Ecuador en una posición de "adoptante" con desafíos de escala. Además, existe una fuga de talento significativa; los investigadores formados en la región emigran debido a la falta de infraestructura de cómputo y ecosistemas de I+D robustos (CENIA, 2025).
- El Paradigma AI4R: La literatura reconoce a la IA como el "quinto paradigma de la investigación" (AI4R), capaz de acelerar el descubrimiento científico mediante el análisis masivo de datos y la generación de hipótesis (Cedeño Meza et al., 2024; Molina & Medina, 2025).

El análisis de estos datos demuestra que la adopción de la IA en la investigación no es opcional. El Marco IA593, al fomentar explícitamente el uso de IA para la investigación (AI4R) y la modernización de infraestructuras (Pilar V), responde a la urgencia de evitar la irrelevancia científica. No se trata solo de usar herramientas para escribir papers, sino de integrar metodologías de machine learning y data science en todas las disciplinas para potenciar la capacidad de descubrimiento endógeno y la soberanía tecnológica de la universidad (Sieker et al., 2025).

### 3.4. Aceleración científica y la "División de GPU"

El AI Index Report 2025 de la Universidad de Stanford confirma que la IA se ha convertido en un motor indispensable para el descubrimiento científico (AI for Science), citando avances como AlphaFold 3 en biología. Sin embargo, el reporte introduce el concepto de la "División de GPU" (The GPU Divide), donde la brecha ya no es solo de acceso a internet, sino de acceso a capacidad de cómputo de alto rendimiento para entrenar o ajustar modelos (Maslej et al., 2025).

Esto valida la urgencia del Pilar III (Investigación) y Pilar V (Infraestructura) del Marco IA593. Las universidades que no inviertan en infraestructura compartida o alianzas de cómputo (como se sugiere en el Atlas de IA) quedarán relegadas a ser meras consumidoras de APIs comerciales, limitando su soberanía investigativa (Maslej et al., 2025; PNUD, 2025b).

**3.5. La convergencia tecnológica y la saturación de benchmarks**

El AI Index Report 2025 señala un punto de inflexión tecnológica: la "convergencia en la frontera". Los modelos de IA líderes (GPT-5, Claude 3, Gemini) han alcanzado niveles de rendimiento casi idénticos y han saturado los benchmarks tradicionales de evaluación (como MMLU o HumanEval), superando la línea base humana en tareas de comprensión lectora y codificación (Maslej et al., 2025; PNUD, 2025b).

Este hallazgo es fundamental para el Pilar II (Docencia) del Marco IA593. Si las herramientas comerciales ya resuelven exámenes estandarizados con precisión experta, la universidad no puede seguir evaluando resultados finales. El marco teórico se alinea con la necesidad de evaluaciones de proceso y pensamiento crítico ("Humanity's Last Exam"), ya que la capacidad técnica de la IA ha convertido la detección de plagio basada en texto en una carrera armamentista obsoleta (Maslej et al., 2025; PNUD, 2025b).

**3.6. Análisis de trabajos relacionados y brechas existentes**

La revisión de la literatura reciente (2024–2025) evidencia un ecosistema de marcos de trabajo disperso y poco articulado, que puede agruparse en tres grandes categorías: i) enfoques de gobernanza ética de alto nivel, ii) propuestas de integración pedagógica de alcance específico y iii) modelos orientados a la gestión de riesgos. Si bien estos enfoques resultan fundamentales para orientar el uso de la inteligencia artificial, su aplicación presenta limitaciones operativas y una débil contextualización institucional, especialmente en entornos de educación superior pública. Frente a estos vacíos, el Marco IA593 se plantea como una respuesta integradora, orientada a articular principios, prácticas y mecanismos de implementación adaptados al contexto universitario.

A. Marcos de gobernanza regulatoria y políticas institucionales: En un primer nivel, la literatura especializada se caracteriza por una amplia presencia de marcos conceptuales orientados a la formulación de políticas institucionales y a la regulación del uso de la inteligencia artificial. (Wang et al., 2024) proponen el Ethical Regulatory Framework for AI (ERF-AI), el cual integra teorías multidisciplinarias para agencias reguladoras, enfocándose en la configuración de roles y procedimientos de control dinámico. En una línea similar, (Azevedo et al., 2025) analizan políticas institucionales en educación superior, destacando la necesidad de una gobernanza compartida y la alfabetización en IA para el profesorado, aunque su enfoque está muy centrado en el contexto estadounidense y el cumplimiento de normativas como FERPA. Por su parte, (Acuña Contreras et al., 2024) presentan una propuesta sólida de gobernanza de datos para universidades, vital para la toma de decisiones, pero limitada al activo de los datos sin abarcar la dimensión generativa de la IA.

   *Brecha identificada*: Estos marcos proporcionan una base teórica robusta ("el qué"), pero carecen de la granularidad operativa para las IES ecuatorianas. No abordan la alineación con el CES ni resuelven la tensión de la "pobreza de GPU" o infraestructura de cómputo identificada en el marco IA593.

B. Modelos de integración pedagógica y alfabetización: En un segundo nivel, existen múltiples propuestas centradas en la enseñanza. (Cacho, 2024) presenta un modelo de directrices equilibradas (enfoque de las "6C": consultar, citar, verificar, etc.) para la integridad académica. (Abbasnejad et al., 2025) desarrollan un marco multinivel específico para la educación en ingeniería, enfocándose en la adaptación curricular y la infraestructura técnica para disciplinas STEM. Asimismo, (Adeoye et al., 2025) proponen un marco integrador para la enseñanza y la evaluación utilizando la teoría de sistemas adaptativos complejos.

   *Brecha identificada*: Si bien estos trabajos abordan eficazmente el pilar de Docencia, tienden a aislar la enseñanza de la función sustantiva de Investigación (AI4R). El Marco IA593 se distingue al establecer protocolos técnicos para la producción científica (como las "Redes de Prompts") y la gestión de la investigación, áreas críticas para evitar la irrelevancia académica en la región.

C. Enfoques basados en riesgos y Human-Centered AI: Finalmente, propuestas recientes como HEAT-AI de (Temper et al., 2025) introducen un enfoque basado en riesgos (mínimo, limitado, alto) inspirado en la Ley de IA de la UE, regulando casos de uso específicos. Simultáneamente, (Le Dinh et al., 2025) proponen un marco de IA centrada en el humano (HCAI) específicamente para revisiones sistemáticas de literatura, asegurando la supervisión humana en procesos de investigación. (González-Fernández et al., 2025) revisan los desafíos éticos y normativos, subrayando la necesidad de códigos éticos académicos.

   *Brecha identificada*: Aunque el enfoque de riesgos es vital, muchos de estos marcos son reactivos. El Marco IA593 llena este vacío mediante su estructura pentagonal (STEER, LEARN, DISCOVER, CONNECT, OPERATE), ofreciendo no solo directrices éticas, sino una hoja de ruta operativa para la infraestructura tecnológica sostenible y la vinculación con la sociedad, aspectos a menudo ausentes en modelos puramente regulatorios.

**Tabla I.** Comparativa del Marco IA593 con trabajos relacionados relevantes.

| **Característica / Dimensión** | **Marco IA593** | **HEAT-AI** (Temper et al., 2025) | **HCAI-SLR** (Le Dinh et al., 2025) | **Marco Ingenieril** (Abbasnejad et al., 2025) | **Políticas Institucionales** (Azevedo et al., 2025) |
|---|---|---|---|---|---|
| **Enfoque Principal** | Sistémico (Gobernanza, Docencia, Investigación, Gestión, Vinculación) | Regulación basada en niveles de riesgo (UE AI Act) | Metodología de investigación (Revisiones Sistemáticas) | Integración técnica y pedagógica en Ingeniería | Políticas de facultad y gobernanza compartida |
| **Alineación Normativa Local** | **Alta** (Específica para Ecuador: CES, LOPDP, EFIA-EC) | Baja (Enfocado en normativa europea/UE) | Baja (Enfoque metodológico general) | Media (Contexto australiano/global) | Media (Contexto estadounidense/FERPA) |
| **Gestión de Infraestructura (GPU)** | **Explícita** (Aborda la "pobreza de cómputo" y sostenibilidad) | No abordado explícitamente | No abordado | Mencionada como soporte técnico | Mencionada superficialmente |
| **Investigación (AI4R)** | **Pilar Central (DISCOVER)**: Protocolos de "Redes de Prompts" y no coautoría | Mencionado como caso de riesgo | **Foco exclusivo**: Procesos de revisión de literatura | Integrado en prácticas disciplinares | Mencionado en términos de integridad |
| **Vinculación con la Sociedad** | **Pilar Central (CONNECT)**: IA para ODS y territorio | No es un foco principal | No aplica | No es un foco principal | No es un foco principal |
| **Operatividad** | **Alta**: KPIs definidos y alineados al PEDI institucional | Alta: Clasificación clara de casos de uso prohibidos/permitidos | Alta: Pasos específicos para procesos de investigación | Media: Observaciones etnográficas y bibliométricas | Media: Recomendaciones de políticas |

Como se observa en la Tabla I, aunque existen marcos robustos para la regulación de riesgos (Temper et al., 2025) o metodologías específicas de investigación (Le Dinh et al., 2025), el Marco IA593 es el único que integra la gestión de infraestructura tecnológica (vital para países en desarrollo) y la vinculación con la sociedad dentro de un modelo de gobernanza alineado a la normativa nacional ecuatoriana. Esto valida su originalidad y pertinencia como instrumento para evitar la dependencia tecnológica y fomentar la soberanía académica.

## 4. Metodología de construcción del Marco IA593

El diseño del Marco de Trabajo IA593 se apoyó inicialmente en el uso estratégico de herramientas de inteligencia artificial generativa como apoyo al análisis, síntesis y organización de un corpus amplio de evidencia científica y normativa. En una primera instancia, estas herramientas permitieron integrar los resultados de una revisión sistemática de la literatura, junto con los marcos regulatorios nacionales e institucionales, para construir un diseño base del marco conceptual. Este diseño preliminar no constituyó un resultado final, sino un insumo analítico que fue posteriormente sometido a un proceso riguroso de depuración y validación (Figura 2). Sobre esta base, el Marco IA593 se consolidó mediante un proceso lógico estructurado en tres etapas secuenciales, orientadas a alinear el mandato normativo ecuatoriano con criterios éticos, estratégicos y operativos, garantizando que la adopción de la IA se configure como un eje transversal que fortalezca, y no simplemente complemente, las funciones sustantivas de la universidad.

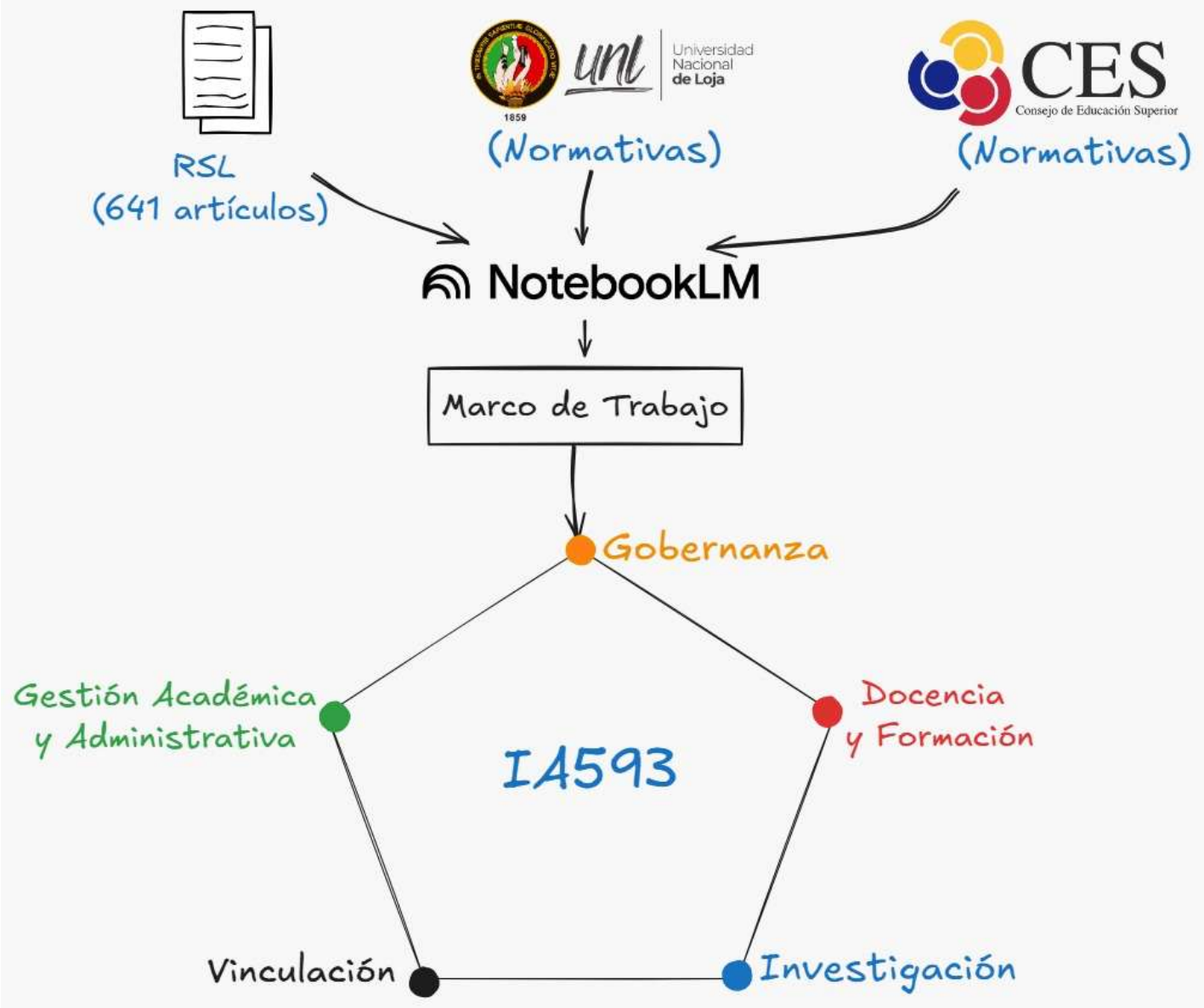

**Figura 2.** Proceso de diseño del Marco de Trabajo IA593 apoyado en herramientas de IA Gen y validado mediante tres etapas metodológicas (generado con Gemini).

### 4.1. Fase 1: Alineación con las funciones sustantivas del sistema de educación superior

La arquitectura base del marco (sus 5 pilares) se diseñó tomando como referencia explícita las Funciones Sustantivas definidas en el artículo 117 de la Ley Orgánica de Educación Superior (LOES) y desarrolladas en el RRA del CES. Se estableció una correspondencia directa entre los mandatos del RRA y los pilares del IA593 para asegurar la validez jurídica y la pertinencia académica del modelo:

- Docencia (Art. 4.a del RRA): El reglamento define la docencia como la construcción de conocimientos y el desarrollo de capacidades. El Marco IA593 responde a esto con el Pilar II (LEARN), enfocándose en la Alfabetización en IA y la actualización curricular para garantizar la formación integral exigida por la norma (CES, 2023).
- Investigación (Art. 4.b del RRA): Definida como una labor creativa, sistemática y sustentada en principios éticos. El marco integra este mandato en el Pilar III (DISCOVER), estableciendo protocolos para el uso de IA para la investigación (AI4R) que aseguren el rigor epistemológico y la integridad científica frente a la generación sintética de contenidos (CES, 2023; Molina & Medina, 2025).
- Vinculación con la sociedad (Art. 4.c del RRA): El mandato de generar capacidades e intercambio de conocimientos con el entorno se operacionaliza en el Pilar IV (CONNECT), orientando el uso de la IA hacia la solución de problemas del territorio y el cumplimiento de los ODS (CES, 2023; IESALC, 2021).
- Gestión institucional: Aunque el RRA regula la gestión de las funciones sustantivas (Art. 2), el marco desglosa esta dimensión en dos pilares de soporte necesarios para la era digital:
  - Pilar I (STEER - Gobernanza): Para la dirección estratégica, normativa y ética (Comités de Ética).
  - Pilar V (OPERATE - Gestión): Para la infraestructura tecnológica y la automatización administrativa.

### 4.2. Fase 2: Integración de principios éticos (enfoque FATE y UNESCO)

Una vez definida la estructura funcional basada en el CES, se aplicó una capa transversal de principios éticos para "blindar" las funciones sustantivas ante los riesgos de la IA. Se seleccionó el modelo FATE (Fairness, Accountability, Transparency, Ethics) y la Recomendación sobre la Ética de la IA de la UNESCO para cualificar cada pilar:

- Transparencia: Se integró como requisito en la docencia (citación de uso de IA) y la investigación (declaración en metodologías).
- Equidad (Fairness): Se aplicó en la gestión para auditar sesgos algorítmicos que pudieran afectar la admisión o evaluación de estudiantes.
- Responsabilidad (Accountability): Se formalizó mediante la creación del Comité de Ética de IA (CEIA) en el pilar de gobernanza (UNESCO, 2024).

### 4.3. Fase 3: Operacionalización estratégica (PEDI UNL)

Finalmente, el marco se contextualizó mediante su alineación con el Plan Estratégico de Desarrollo Institucional (PEDI 2024-2028). Cada pilar del IA593 se vinculó a una meta concreta del PEDI (ej. Meta OT6 para investigación, Meta OT8 para infraestructura tecnológica), transformando el marco teórico en un instrumento de gestión medible con indicadores de desempeño (KPIs) claros (Universidad Nacional de Loja, 2025).

Esta metodología garantiza que el IA593 no sea una imposición tecnológica externa, sino la evolución natural de las funciones sustantivas de la universidad (Docencia, Investigación, Vinculación) adaptadas a la realidad algorítmica bajo la normativa del CES.

## 5. Marco de Trabajo IA593

El Marco de Trabajo IA593 se constituye como el instrumento rector para la adopción de la IA en la universidad. Su arquitectura sistémica se basa en los principios éticos globales FATE (Equidad, Rendición de Cuentas, Transparencia y Ética) y en la Recomendación sobre la Ética de la IA de la UNESCO, adaptándolos a las brechas estructurales y oportunidades del contexto ecuatoriano identificadas en el índice ILIA 2025 (CENIA, 2025), asegurando que la tecnología sirva como un amplificador de las capacidades humanas y no como un sustituto no supervisado (Giandana Gigena et al., 2024; UNESCO, 2024).

El Marco IA593 se estructura en cinco pilares interdependientes que cubren la totalidad de las funciones sustantivas y de gestión institucional. A continuación, se detalla la composición exhaustiva de cada pilar.

#### 5.1.1. Pilar I: Gobernanza Transversal y Ética Institucional (STEER)

Este pilar actúa como el "cerebro" del sistema, estableciendo la dirección estratégica y la supervisión normativa. Ante la ausencia de una Ley de IA específica en Ecuador, la universidad asume la autorregulación basándose en la responsabilidad proactiva (Tabla II).

- **Fundamento:** Se establece la obligatoriedad de realizar evaluaciones de impacto ético (EIE) para sistemas de alto riesgo (ej. algoritmos de admisión o vigilancia), conforme a los estándares de la OCDE y la estrategia nacional de IA (MINTEL, 2026; OECD, 2025).

**Tabla II.** Pilar I: Gobernanza Transversal y Ética Institucional (STEER) del Marco IA593.

| Componente | Descripción Operativa | Ámbito (Alcance) | Actores Responsables | Indicadores (KPIs) y Metas PEDI | Fundamentos científicos y teóricos |
|---|---|---|---|---|---|
| Gobernanza Estratégica y Comité de Ética (CEIA) | Consolidación de la hoja de ruta de IA alineada a la visión institucional. Establecimiento formal del Comité de Ética de IA (CEIA) para supervisar sistemas de alto riesgo mediante Evaluaciones de Impacto Ético (EIE) previas a su despliegue. | Gestión, Docencia, Investigación, Vinculación | Estratégico: Rectorado, Vicerrectorado.<br><br>Gestión: Dirección de Planificación (DPD), CEIA. | • Constitución oficial del CEIA.<br>• % de sistemas de alto riesgo con EIE aprobada.<br>• Actualización de normativa de investigación (Meta OT6). | • PEDI UNL 2024-2028 (Universidad Nacional de Loja, 2025)<br>• (Mahrishi et al., 2024)<br>• (Wang et al., 2024)<br>• (George & Wooden, 2023) |
| Protección de Datos y Gestión de Riesgos | Protocolos para garantizar la privacidad, anonimización y seguridad de datos (biométricos, académicos). Incluye el reporte obligatorio de desviaciones o sesgos al CEISH-UNL y el cumplimiento de la LOPDP. | Gestión, Investigación | Técnico: Dirección de TI (DTI), Oficial de Protección de Datos.<br><br>Supervisión: CEIA/CEISH-UNL. | • Reporte anual de auditorías algorítmicas.<br>• Tasa de reducción de incidentes de privacidad.<br>• Cumplimiento de protocolos de ciberseguridad. | • (Hasan et al., 2024)<br>• (Acuña Contreras et al., 2024)<br>• CEISH-UNL (Comité de Ética de Investigación en Seres Humanos de la Universidad Nacional de Loja, 2023) |

### 5.2. Pilar II: Docencia y Formación (LEARN)

Enfocado en la actualización de las competencias humanas. Rechaza la prohibición de la IA y promueve una Alfabetización en IA (AI Literacy) crítica, donde docentes y estudiantes entienden no solo cómo usar las herramientas, sino cómo funcionan y cuáles son sus implicaciones éticas (Tabla III).

- **Fundamento:** Adopción del Marco de Competencias en IA para Estudiantes y Docentes de la UNESCO, priorizando el pensamiento crítico y la “ética encarnada” sobre la mera destreza técnica. Se exige la transparencia en el uso de IA generativa en trabajos académicos para preservar la integridad (UNESCO, 2025d, 2025c).

**Tabla III.** Pilar II: Docencia y Formación (LEARN) del Marco IA593.

| Componente | Descripción Operativa | Ámbito (Alcance) | Actores Responsables | Indicadores (KPIs) y Metas PEDI | Fundamentos científicos y teóricos |
|---|---|---|---|---|---|
| Alfabetización Integral en IA (AI Literacy) | Programa transversal obligatorio para docentes y estudiantes que cubre uso técnico (prompts), ético y crítico de la IA, basado en el Marco de Competencias de la UNESCO. | Docencia, Gestión | Académico: Vicerrectorado Académico (VA), UEDL.<br><br>Ejecución: Docentes. | • % de docentes certificados en competencias de IA.<br>• % de estudiantes que aprueban módulos de alfabetización digital. | • (UNESCO, 2025c)<br>• (Burneo-Arteaga et al., 2025)<br>• (Southworth et al., 2023) |
| Integridad Académica y Evaluación Crítica | Rediseño de evaluaciones para priorizar el juicio humano sobre la generación automática. Exigencia de declarar explícitamente el uso de IA en trabajos académicos (citación y transparencia). | Docencia | Control: Docentes, VA.<br><br>Usuarios: Estudiantes. | • Inclusión de la “Cláusula de IA” en el 100% de los sílabos.<br>• Tasa de incidentes de fraude académico detectados.<br>• % de evaluaciones rediseñadas. | • (An et al., 2025)<br>• (Bittle & El-Gayar, 2025)<br>• (Guizani et al., 2025)<br>• (Cacho, 2024) |
| Personalización del Aprendizaje | Implementación de tutores inteligentes y sistemas de retroalimentación automatizada, siempre bajo supervisión docente (“human-in-the-loop”) para evitar la dependencia tecnológica. | Docencia | Académico: Docentes.<br><br>Tecnológico: UEDL/TI. | • % de cursos que utilizan asistentes virtuales de apoyo.<br>• Satisfacción estudiantil con herramientas personalizadas. | • (Kakhkharova & Tuychieva, 2024) |

## 5.3. Pilar III: Investigación (DISCOVER)

Reconoce a la IA como el quinto paradigma de investigación (AI4R), potenciando el descubrimiento científico mediante el análisis masivo de datos, pero bajo estrictos protocolos de supervisión ética (Tabla IV).

- **Fundamento:** La IA no puede ser coautora de publicaciones científicas (COPE/Nature/Elsevier). El marco establece que el investigador es el único responsable (accountable) de los resultados. Se protege la privacidad de los datos sensibles de sujetos de investigación reportando al CEISH-UNL (Cedeño Meza et al., 2024; Molina & Medina, 2025).

### 5.3.1. Protocolo de trazabilidad metodológica: Redes de prompts y niveles de supervisión

Para operacionalizar la transparencia en la producción científica de la UNL, se establece el siguiente protocolo de reporte obligatorio para trabajos de titulación de grado/posgrado y artículos, basado en la taxonomía de interacción humano-IA:

1) Definición de la interacción (Prompt Engineering Documentado): No basta con citar la herramienta; el investigador debe documentar la “Red de Prompts” (Prompt Net) (Penabad-Camacho et al., 2024). Se requiere diferenciar entre:
    a. Prompt Inicial: La instrucción base dada al modelo.
    b. Prompts Secundarios: Instrucciones de refinamiento, enriquecimiento o delimitación para mitigar alucinaciones.
    c. Prompts Complementarios: Instrucciones para formato, traducción o corrección de estilo.
2) Diagrama de flujo de decisión: Todo trabajo de investigación que utilice IA Generativa debe incluir en su metodología un diagrama de flujo que visualice en qué secciones (Introducción, Metodología, Resultados) se utilizó la IA y bajo qué nivel de supervisión. Se debe declarar explícitamente si hubo un uso filtrado (verificado por humano) o si se descartó material por sesgo (Penabad-Camacho et al., 2024).
3) Clasificación de herramientas por propósito (Taxonomía Funcional): Siguiendo la bitácora de herramientas digitales, los investigadores deben clasificar el uso de la IA en una de las siguientes categorías funcionales para facilitar la auditoría ética (Hinojosa Mamani et al., 2024):
    a. Búsqueda y mapeo de literatura: (ej. Litmaps, Research Rabbit) para la conexión de conceptos, no para la redacción final.
    b. Análisis y extracción de datos: (ej. SciSpace, Scholarcy) para la síntesis de documentos, requiriendo validación de fuentes originales.
    c. Asistencia de escritura y estilo: (ej. ChatGPT, Rytr) para corrección gramatical o traducción, manteniendo la autoría intelectual humana.
4) Validación de sesgos y datos: En concordancia con las directrices de gobernanza de datos (MTDFP, 2025), se debe reportar si se utilizaron datos personales o sensibles para entrenar o consultar modelos (fine-tuning), garantizando el cumplimiento de la LOPDP mencionada en el Pilar I. Se exige una declaración de que los prompts utilizados no inducen sesgos discriminatorios ni alucinaciones factuales (Penabad-Camacho et al., 2024).

**Tabla IV.** Pilar III: Investigación (DISCOVER) del Marco IA593.

| **Componente** | **Descripción Operativa** | **Ámbito (Alcance)** | **Actores Responsables** | **Indicadores (KPIs) y Metas PEDI** | **Fundamentos científicos y teóricos** |
|---|---|---|---|---|---|
| Integridad Científica y Supervisión Ética | Normativa que prohíbe la coautoría de la IA. Exige transparencia metodológica sobre las herramientas usadas y supervisión del CEISH en proyectos con datos sensibles. | Investigación | Supervisión: Dir. Investigación, CEISH-UNL, Comités Editoriales.<br><br>Ejecución: Investigadores. | • Protocolos con IA aprobados por CEISH.<br>• Tasa de declaración de herramientas en *papers*.<br>• Cumplimiento ético en manejo de *datasets*. | • (Masbernat, A.C. et al., 2024)<br>• (McLennan et al., 2020)<br>• (P. Rivas et al., 2023) |
| IA para la Investigación (AI4R) | Fomento del uso de *Machine Learning* y procesamiento de lenguaje natural (NLP) para el análisis masivo de datos, modelos predictivos y revisión de literatura en todas las disciplinas. | Investigación | Gestión: Dir. Investigación.<br><br>Ejecución: Grupos de Investigación. | • Índice de producción académica per cápita (Meta I5).<br>• Proyectos de investigación pertinentes ejecutados (Meta OT7). | • (Yu et al., 2024)<br>• (Ramírez Téllez et al., 2024) |

## 5.4. Pilar IV: Vinculación y Extensión (CONNECT)

Orienta el poder de la IA hacia la resolución de problemas sociales y productivos del entorno, evitando que la tecnología aumente las brechas de desigualdad en el territorio (Tabla VI).

- **Fundamento:** Alineación con los ODS. La universidad actúa como un nodo de transferencia tecnológica hacia el sector productivo y la sociedad civil, fomentando una "IA para el bien" (AI for Good) (CENIA, 2025; Fábrega Lacoa et al., 2024).

**Tabla VI.** Pilar IV: Vinculación y Extensión (CONNECT) del Marco IA593.

| **Componente** | **Descripción Operativa** | **Ámbito (Alcance)** | **Actores Responsables** | **Indicadores (KPIs) y Metas PEDI** | **Fundamentos científicos y teóricos** |
|---|---|---|---|---|---|
| IA para el Desarrollo Sostenible (ODS) | Desarrollo de proyectos de vinculación que apliquen IA para resolver desafíos locales (ej. agro, salud, ambiente) y reducir la brecha digital en comunidades vulnerables. | Vinculación | Gestión: Dir. Vinculación.<br><br>Ejecución: Docentes, Estudiantes. | • Número de proyectos de vinculación derivados de I+D (Meta I2).<br>• Cantidad de beneficiarios en programas de reducción de brecha digital. | • (Li et al., 2025)<br>• (Salas et al., 2022) |
| Conexión Industria-Academia | Alianzas estratégicas con el sector productivo para transferencia tecnológica y actualización de perfiles profesionales según la demanda laboral de la era de la IA. | Vinculación, Docencia | Gestión: Dir. Vinculación, VA.<br><br>Aliados: Empresas/Cámaras. | • Número de convenios activos para IA aplicada.<br>• Eventos de educación continua ejecutados (Meta I3). | • (Castelló-Sirvent et al., 2024)<br>• (Pradit & Piriyasurawong, 2024)<br>• (OECD, 2023b) |

### 5.5. Pilar V: Gestión Académica y Administrativa (OPERATE)

Provee el soporte operativo necesario para que los otros pilares funcionen: infraestructura de cómputo, datos de calidad y procesos administrativos eficientes (Tabla VII).

- **Fundamento:** La "Gestión Institucional Inteligente" del PEDI UNL requiere modernizar la infraestructura tecnológica para soportar cargas de trabajo de IA y automatizar procesos burocráticos repetitivos, liberando talento humano para tareas de valor (Universidad Nacional de Loja, 2025).

**Tabla VII.** Pilar V: Gestión Académica y Administrativa (OPERATE) del Marco IA593.

| **Componente** | **Descripción Operativa** | **Ámbito (Alcance)** | **Actores Responsables** | **Indicadores (KPIs) y Metas PEDI** | **Fundamentos científicos y teóricos** |
|---|---|---|---|---|---|
| Infraestructura Tecnológica Sostenible | Inversión en hardware (GPUs, servidores), conectividad y plataformas interoperables necesarias para soportar el cómputo de IA, asegurando sostenibilidad financiera. | Gestión | Infraestructura: DTI, Dir. Administrativa, Dir. Desarrollo Físico. | • % de implementación de infraestructura tecnológica.<br>• Escenarios modernizados (Meta OT8). | • PEDI UNL 2024-2028 (Universidad Nacional de Loja, 2025)<br>• (Elhajji et al., 2020) |
| Gestión Inteligente y Automatización | Automatización de procesos administrativos rutinarios (RPA) y uso de analítica de datos para la toma de decisiones basada en evidencia (gestión institucional inteligente). | Gestión | Estratégico: DPD, Coord. Administrativa.<br><br>Técnico: DTI. | • % de unidades que implementan gestión por procesos automatizados.<br>• Reducción de carga administrativa manual. | • (Ramírez Téllez et al., 2024)<br>• (Ruiz Muñoz et al., 2024) |
| Capacitación del Talento Humano (Admin/TI) | Programa de *reskilling* y *upskilling* digital para el personal administrativo y de TI, enfocado en la gestión de nuevos sistemas y seguridad de la información. | Gestión | Gestión: Talento Humano, DTI. | • Número de programas de capacitación en IA para administrativos.<br>• % de personal de TI certificado en mantenimiento de sistemas IA. | • (Iskandarova et al., 2024) |

## 6. Guía de implementación y alineación normativa

La adopción del Marco IA593 no debe interpretarse como una carga regulatoria adicional, sino como un mecanismo de aseguramiento de la calidad que garantiza el cumplimiento de la normativa nacional vigente ante la disrupción tecnológica. Para facilitar su replicabilidad en otras universidades públicas y privadas del Ecuador, se presenta a continuación la alineación jurídica explícita y una hoja de ruta de despliegue progresivo.

### 6.1. Matriz de correspondencia normativa (CES - IA593)

La matriz vincula los componentes operativos del Marco IA593 con los artículos mandatorios del RRA del CES y la LOPDP, proporcionando la seguridad jurídica necesaria para la toma de decisiones institucionales (Tabla VIII).

**Tabla VIII.** Correspondencia entre normativa y componentes del Marco IA593.

| **Componente IA593** | **Normativa Vigente (Base Legal)** | **Acción Sugerida para la IES** |
|---|---|---|
| Integridad Académica y Evaluación Crítica (Pilar II) | RRA - Art. 38 (Fraude y Deshonestidad Académica): Tipifica el fraude como la apropiación de ideas o el uso de soportes no autorizados en procesos de evaluación (CES, 2023). | Actualizar el Reglamento de Ética y Disciplina: Incluir explícitamente el “uso no declarado de IA Generativa” como una tipología de fraude académico, diferenciándolo del uso asistido autorizado (Universidad Andrés Bello, 2024). |
| Integridad Científica y AI4R (Pilar III) | RRA - Art. 34 (Investigación): Define la investigación como una labor creativa y sistemática bajo principios éticos. Normas COPE/UNESCO: La IA no puede ser autora (CES, 2023; UNESCO, 2024). | Protocolo de Publicación: Exigir en las normativas de titulación y revistas institucionales la “Declaración de Uso de IA” en la metodología, prohibiendo la coautoría de la IA (Cedeño Meza et al., 2024). |
| Alfabetización en IA (Pilar II) | RRA - Art. 4 (Funciones Sustantivas): Obligación de garantizar una formación integral y el desarrollo de capacidades profesionales (CES, 2023). | Capacitación Docente: Incorporar un módulo obligatorio de *AI Literacy* (basado en el Marco UNESCO) en el plan de formación continua del profesorado (UNESCO, 2025d). |
| Protección de Datos (Pilar I y V) | LOPDP - Art. 42: Exige Evaluaciones de Impacto para el tratamiento de datos de alto riesgo (como el perfilamiento de estudiantes) (MINTEL, 2026). | Auditoría de Herramientas: Implementar un procedimiento de validación técnica previo a la adquisición de software de IA, verificando que los servidores cumplan con la normativa de privacidad local (The Institute of Internal Auditors, 2023). |
| Infraestructura Tecnológica (Pilar V) | RRA - Art. 61 (Recursos de Aprendizaje): Exige garantizar la disponibilidad de recursos tecnológicos y bibliográficos pertinentes para el aprendizaje (CES, 2023). | Modernización de Laboratorios: Inversión en infraestructura de cómputo (GPU/Nube) y acceso a bibliotecas digitales con asistentes de investigación inteligentes (ej. Semantic Scholar) (Molina & Medina, 2025). |

### 6.2. Fases de despliegue institucional

Se propone una implementación escalonada en tres fases, utilizando la metodología RAM (Readiness Assessment Methodology) de la UNESCO como punto de partida.

**6.2.1. Fase 1. Diagnóstico y gobernanza (meses 1-3):** El objetivo es entender la "madurez de IA" de la institución y establecer las reglas del juego.

- Acción 1 (Diagnóstico RAM): Aplicar la *Metodología de Evaluación del Estadio de Preparación (RAM)* de la UNESCO para identificar brechas en infraestructura, datos y gobernanza ética dentro de la universidad (UNESCO, 2025e, 2025a).
- Acción 2 (Comité CEIA): Constituir el Comité de Ética de IA (CEIA) o asignar sus funciones al Comité de Ética existente, integrando perfiles técnicos y jurídicos (OdiseIA, 2022).
- *Entregable:* Informe de Diagnóstico Institucional y Reglamento del CEIA aprobado.

**6.2.2. Fase 2. Piloto y alfabetización (meses 4-9):** Enfocada en la capacitación ("aprender haciendo") y la actualización curricular sin cambios normativos drásticos aún.

- Acción 1 (Formación de Formadores): Ejecutar un programa piloto de alfabetización para un grupo núcleo de docentes innovadores, enfocándose en ingeniería de prompts y rediseño de evaluaciones (Universidad Andrés Bello, 2024; Villegas Dianta et al., 2025).
- Acción 2 (Guías Orientativas): Publicar guías de "Buenas Prácticas" para estudiantes y docentes que clarifiquen qué usos están permitidos (ej. lluvia de ideas, corrección de estilo) y cuáles prohibidos (ej. generación total de texto sin revisión) (Grupo de trabajo de Inteligencia Artificial Generativa de la UNAM, 2025; Serrano Acitores, 2025).
- *Entregable:* 20% de la planta docente certificada y Guía de Uso Ético publicada.

**6.2.3. Fase 3. Institucionalización y normativa (meses 10-12):** Formalización de las políticas en los estatutos y sistemas de gestión de la universidad.

- Acción 1 (Actualización Reglamentaria): Modificar el Reglamento de Régimen Académico interno y el Reglamento de Titulación para incluir las cláusulas de integridad y uso de IA definidas en la Matriz de Correspondencia (Sección 6.1, Tabla VIII).
- Acción 2 (Infraestructura AI4R): Desplegar herramientas institucionales de IA para investigación (ej. Scite, Elicit) y gestión, asegurando licencias corporativas que protejan los datos (Cedeño Meza et al., 2024).
- *Entregable:* Normativa interna reformada y ecosistema tecnológico operativo.

## 7. Proyección estratégica 2026-2030: Hacia la madurez institucional y la IA agéntica

La rápida evolución tecnológica, marcada por las predicciones para 2026, exige que el Marco IA593 no sea estático. Según el Informe de Predicciones Tecnológicas 2026, la ventaja competitiva de las instituciones se desplazará de la mera escala de personal al "apalancamiento de la inteligencia" mediante agentes de IA que actuarán como miembros estándar de los equipos de trabajo (IEEE Computer Society, 2026). Para evitar que la UNL y el sistema ecuatoriano queden rezagados en una "velocidad de adopción técnica" sin impacto social real (Cruz et al., 2025), se establecen cuatro ejes de evolución estratégica.

### 7.1. De la IA generativa a la IA agéntica en la educación

Si bien la fase inicial del Marco IA593 se centró en la alfabetización sobre modelos de lenguaje (LLM), el horizonte 2026 demanda la integración de Sistemas de Tutoría Inteligente (ITS) y agentes autónomos. La OCDE señala que la IA generativa está en transición hacia herramientas capaces de ofrecer andamiaje cognitivo y retroalimentación personalizada en tiempo real, superando la función de simples generadores de texto (OECD, 2026).

- Implementación de agentes de IA: Se debe transitar del uso de chatbots pasivos a "agentes de IA" capaces de planificar y ejecutar secuencias de tareas complejas en los flujos de trabajo institucionales (OECD, 2026). Esto implica redefinir el rol docente y administrativo, donde la competencia clave será la gestión y supervisión de estos agentes para evitar la dependencia tecnológica y la deuda cognitiva (OECD, 2026).
- Campus AI-Ready: Siguiendo las directrices de infraestructura para 2026, la UNL debe evolucionar hacia un "Campus Preparado para IA", donde la conectividad y el hardware soporten no solo el consumo de datos, sino la operación segura de estos agentes en el borde (edge computing) (Microsoft, 2025).

### 7.2. Soberanía científica y metodologías rigurosas (AI4S)

El estado de la ciencia en 2025 evidencia una convergencia estratégica entre la computación de alto rendimiento, la IA y, de manera incipiente, la computación cuántica, lo que está redefiniendo los procesos de generación, análisis y validación del conocimiento científico (OEI-UNESCO, 2025). En este escenario, caracterizado por el uso intensivo de datos, modelos computacionales complejos y herramientas automatizadas, resulta indispensable que la investigación desarrollada en la UNL adopte criterios explícitos y homogéneos para documentar sus decisiones metodológicas, los datos utilizados, los procesos de análisis y los apoyos computacionales empleados. Esta estandarización del reporte metodológico es una condición clave para garantizar la reproducibilidad, la comparabilidad de resultados y la integridad ética de la investigación, en línea con las prácticas internacionales descritas en los principales indicadores de ciencia y tecnología.

- Protocolo de "Prompt Net": Para mitigar la crisis de reproducibilidad y las "alucinaciones" en la literatura científica, se adopta el protocolo de Red de Prompts (Prompt Net). Los investigadores deben diagramar y declarar no solo el prompt inicial, sino los prompts de refinamiento, enriquecimiento y delimitación utilizados en cada sección del artículo (Introducción, Metodología, Resultados), garantizando la transparencia del proceso (Penabad-Camacho et al., 2024).
- Investigación bibliográfica aumentada: El uso de herramientas como Scite o Elicit debe ser obligatorio para verificar la veracidad de las citas, dado el alto riesgo de referencias fabricadas por modelos no supervisados. La IA debe usarse para mapeo de conocimiento y no como sustituto del juicio crítico autoral (Hinojosa Mamani et al., 2024).

### 7.3. Gobernanza de datos y entornos de pruebas (Sandboxes)

La región de América Latina y el Caribe enfrenta el riesgo de una adopción fragmentada donde las PYMES y universidades pequeñas carecen de estructuras de gobernanza robustas (Cruz et al., 2025). Para contrarrestarlo, el Marco IA593 instituye el concepto de Sandbox Académico de IA.

- Entorno controlado de pruebas: Siguiendo las recomendaciones para la implementación de sandboxes de IA (Guridi & Trivelli, 2025), la UNL creará espacios controlados para probar algoritmos de alto riesgo (ej. admisión o evaluación docente) antes de su despliegue general. Esto permitirá identificar y corregir sesgos algorítmicos y fallos de cumplimiento normativo en un entorno seguro (MTDFP, 2025).
- Gestión pública inteligente: La administración universitaria de la UNL adoptará herramientas de IA para el diseño y evaluación de programas, siguiendo las pautas de modernización del Estado, pero manteniendo siempre una supervisión humana significativa en decisiones que afecten derechos estudiantiles o laborales (Naranjo Bautista et al., 2025).

### 7.4. Ética aplicada y resistencia al determinismo tecnológico

Frente a la narrativa de que la IA es "inevitable", la UNL debe mantener una postura de resistencia al determinismo. Las guías institucionales no deben ser meros manuales de uso, sino instrumentos que cuestionen las implicaciones laborales y ecológicas de la tecnología (Guersenzvaig & Monett, 2026).
- Auditoría de impacto humano y ambiental: Se integrarán métricas de sostenibilidad para evaluar el consumo energético de los modelos utilizados (IEEE Computer Society, 2026) y se vigilarán los impactos psicosociales en la comunidad, como la antropomorfización excesiva de la IA o la pérdida de agencia humana (Guersenzvaig & Monett, 2026).
- Marco de competencias éticas: Se adoptará el marco ampliado de competencias de la UNESCO, que exige que tanto estudiantes como docentes comprendan no solo la técnica, sino la ética de los datos, la privacidad y el impacto social de la IA (UNESCO, 2025b).

### 7.5. Integración regional y competitividad económica

Ecuador, y la región en general, se encuentra en un punto de inflexión donde la IA puede acelerar el desarrollo o exacerbar las brechas (WEF, 2026).
- Colaboración interuniversitaria: Para superar las limitaciones presupuestarias y de talento, se fomentará la creación de consorcios regionales para compartir infraestructura de cómputo y datos anonimizados, siguiendo las recomendaciones para evitar la "fuga de cerebros" y la dependencia tecnológica (Grupo sobre Inteligencia Artificial en la Educación Superior, 2025; Rivas, 2025b).
- Alineación con el mercado laboral 2026: Los currículos se actualizarán para incluir habilidades de alta demanda pronosticadas para 2026, como la ciberseguridad en entornos de IA, la biotecnología asistida por IA y el manejo de asistentes físicos (robótica) y virtuales (IEEE Computer Society, 2026; Microsoft, 2025).

## 8. Discusión

La implementación del Marco IA593 no se produce en un entorno neutral ni aislado, sino que se inserta en un ecosistema universitario complejo, marcado por tensiones financieras, culturales y tecnológicas que condicionan su adopción, apropiación y sostenibilidad institucional. A continuación, se discuten las implicaciones de adoptar este modelo, contrastando los nudos críticos de infraestructura y cultura organizacional con las oportunidades de posicionamiento global.

### 8.1. Desafíos estructurales y culturales

La transición hacia una "Universidad aumentada por IA" enfrenta tres barreras principales que requieren gestión de riesgos activa:

A. La brecha de infraestructura y la pobreza de cómputo (The GPU Divide): El desafío más tangible es el costo del cómputo. A nivel global, se observa una división entre instituciones "ricas en GPU" y "pobres en GPU". El entrenamiento y ajuste (fine-tuning) de modelos de lenguaje requiere una infraestructura de alto rendimiento que es escasa en la región (OECD, 2025). Además, la implementación del Marco IA593 enfrenta la barrera de la "pobreza de GPU" descrita en el Atlas de IA y el reporte de Stanford 2025. El mundo se está dividiendo entre instituciones "ricas en cómputo" (capaces de entrenar modelos) y "pobres en cómputo" (limitadas a consumir APIs) (Sieker et al., 2025).
    - Diagnóstico local: Según la Estrategia Nacional de IA (2026), Ecuador posee una capacidad de GPU de 26.0 Teraflops por millón de habitantes, cifra inferior al promedio regional de 34.2, lo que limita la capacidad de desarrollar modelos soberanos y obliga a depender de APIs comerciales costosas (MINTEL, 2026).
    - Implicación para el marco: El Pilar V (Gestión) del IA593 debe priorizar alianzas con centros de supercómputo regionales (como CEDIA en Ecuador) o negociar licencias corporativas, ya que la inversión en data centers propios enfrenta costos energéticos y de mantenimiento que crecen exponencialmente (PNUD, 2025b), además, para evitar que la UNL sea un mero consumidor pasivo, el Pilar V (Gestión) debe priorizar alianzas con infraestructuras nacionales (como CEDIA) y explorar modelos de lenguaje pequeños (SLMs) que ofrecen rendimiento eficiente con menor costo computacional, una tendencia destacada en el AI Index 2025 como vía para la democratización (Maslej et al., 2025).

B. Resistencia al cambio docente y alfabetización asimétrica: Existe una tensión entre la velocidad de adopción del estudiantado y la curva de aprendizaje del profesorado. Estudios recientes indican que mientras el 63% de las instituciones incentiva el uso de IA, menos de la mitad aborda preocupaciones sobre propiedad intelectual o privacidad, generando incertidumbre docente (Molina & Medina, 2025).
    - El riesgo pedagógico: La resistencia no surge solo del desconocimiento técnico, sino del temor al desplazamiento laboral y a la pérdida de control sobre la evaluación. Sin una alfabetización crítica (Pilar II), existe el riesgo de caer en un "solucionismo tecnológico" donde se adoptan herramientas sin criterio pedagógico, o por el contrario, en una prohibición ineficaz que fomenta el "mercado negro" de tareas automatizadas (Rivas, 2025a; Villegas Dianta et al., 2025).

C. Sesgos algorítmicos y colonización de datos: La mayoría de los modelos grandes de lenguaje (LLM) han sido entrenados con datos del Norte Global, lo que introduce sesgos culturales y lingüísticos que pueden marginar las realidades locales (Pedreño Muñoz et al., 2024).
    - Alucinaciones y veracidad: Los modelos generativos presentan tasas de "alucinaciones" (información falsa presentada con convicción) que desafían la integridad científica. El uso acrítico de estas herramientas en la investigación (Pilar III) sin la supervisión humana experta (human-in-the-loop) puede contaminar la producción científica con referencias inexistentes o datos sesgados (Universidad Pontificia Comillas, 2024; Vivas Urias & Ruiz Rosillo, 2025).

### 8.2. La IA como palanca de diferenciación

A pesar de los desafíos, la adopción temprana y regulada del Marco IA593 ofrece ventajas competitivas invaluables para la universidad en el contexto latinoamericano:

A. Posicionamiento en rankings de innovación y calidad: La capacidad de una universidad para integrar IA en sus procesos es un nuevo indicador de calidad. Instituciones que lideran la adopción de IA, como el Tecnológico de Monterrey o la Universidad de São Paulo, han mejorado su visibilidad y atracción de fondos (Molina & Medina, 2025).
   - Impacto del Pilar III (AI4R): Al potenciar la investigación con herramientas de IA para revisión de literatura masiva y análisis de datos (ej. Semantic Scholar, Scite), la universidad puede incrementar significativamente su output científico y el impacto de sus citas, superando la marginalidad del 0.21% de patentes que actualmente ostenta la región (CENIA, 2025; Molina & Medina, 2025).

B. Atracción y retención de talento humano (fuga de cerebros): América Latina sufre una fuga crónica de talento humano avanzado hacia ecosistemas con mejor infraestructura. El ILIA 2025 destaca que la falta de acceso a cómputo es un factor de expulsión clave (CENIA, 2025).
   - La Universidad como Hub: Al establecer un marco claro (IA593) que garantice acceso a herramientas, seguridad jurídica y soporte ético, la universidad se convierte en un "puerto seguro" para investigadores innovadores que buscan aplicar IA sin riesgos legales. Además, ofrecer una formación curricular actualizada en IA (Pilar II) aumenta la empleabilidad de los egresados, respondiendo a la demanda del mercado laboral que exige competencias digitales avanzadas (CENIA, 2024; Pedreño Muñoz et al., 2024).

C. Liderazgo en responsabilidad social universitaria: Ante la falta de regulación nacional detallada, la universidad tiene la oportunidad de actuar como un "faro ético". Al implementar Comités de Ética de IA (CEIA) y protocolos de protección de datos, la institución no solo cumple la ley, sino que genera confianza pública, posicionándose como un actor clave en la gobernanza tecnológica nacional (MINTEL, 2026).

### 8.3. Costos ocultos y sostenibilidad ambiental

La implementación de la IA en la universidad conlleva un costo ambiental significativo. Según el reporte Estado de la IA en 2025, el entrenamiento y la inferencia de modelos masivos generan una huella de carbono y un consumo hídrico que entran en tensión con las metas de sostenibilidad institucional (OECD, 2025; PNUD, 2025b). El Marco IA593 debe considerar en su Pilar IV (Vinculación) y Pilar V (Gestión) la adopción de "Green AI" o algoritmos eficientes energéticamente.

### 8.4. De chatbots a agentes (Agentic AI)

El estado del arte en 2025 se ha movido de los chatbots pasivos a los agentes de IA (Agentic AI), sistemas capaces de planificar y ejecutar secuencias de tareas complejas (OECD, 2025). Esto abre la oportunidad para que la universidad desarrolle "Agentes Institucionales" que no solo respondan preguntas, sino que gestionen trámites administrativos o asistan en la revisión de literatura de forma autónoma pero supervisada, aumentando la productividad administrativa y científica (Maslej et al., 2025).

### 8.5. Liderazgo en ética y datos locales

A pesar de las brechas técnicas, el ILIA 2025 destaca que América Latina tiene una oportunidad única en el código abierto y la gobernanza de datos locales. Mientras los modelos globales carecen de contexto regional, el Marco IA593 habilita a la universidad para crear datasets propios y curados (patrimonio, biodiversidad, normativa local), contribuyendo a una "IA Soberana" que mitigue los sesgos culturales de los modelos entrenados en el norte global (CENIA, 2025; Pedreño Muñoz et al., 2024).

## 9. Conclusiones y trabajos futuros

La adopción de la IA en la educación superior ecuatoriana ha dejado de ser una ventaja competitiva opcional para convertirse en un requisito de sostenibilidad institucional y pertinencia social. El análisis realizado a lo largo de este trabajo permite establecer las siguientes conclusiones determinantes y proyectar las líneas de desarrollo tecnológico futuro.

### 9.1. De la reacción a la estrategia nacional

- La inevitabilidad operativa: La implementación del Marco de Trabajo IA593 no constituye una elección discrecional, sino la única respuesta operativa viable ante la Estrategia Nacional para el Fomento de la IA (EFIA-EC). Mientras el Estado dicta los lineamientos de gobernanza y ética a nivel macro (Registro Oficial Nº 206), las universidades carecen de los instrumentos "micro" para gestionar el aula y el laboratorio. El IA593 llena este vacío regulatorio, traduciendo principios abstractos en protocolos ejecutables de integridad académica y gestión de datos (MINTEL, 2026).
- Soberanía epistemológica: Frente a un escenario global donde América Latina produce menos del 2% de la investigación en IA, el marco permite transitar de un consumo pasivo de tecnologías del Norte Global a una adopción crítica y soberana. Al regular la AI for Research (AI4R) y exigir transparencia en los algoritmos, la universidad protege su producción científica de la dependencia tecnológica y los sesgos cognitivos inherentes a los modelos comerciales (CENIA, 2025; UNESCO, 2023b).
- Garantía de calidad educativa: La integración de la alfabetización en IA (Pilar II) y la supervisión ética (Pilar I) son mecanismos de aseguramiento de la calidad alineados con el RRA del CES. Ignorar la regulación de la IA no protege la tradición académica, sino que expone a la institución a riesgos legales por violación de la LOPDP y a la devaluación de sus títulos por fraude académico no detectado (CES, 2023; Villegas Dianta et al., 2025).

### 9.2. Trabajos futuros

La aprobación del Marco IA593 habilita una segunda fase de desarrollo tecnológico institucional, enfocada en la personalización y la automatización supervisada:

A. Desarrollo de agentes de IA institucionales (Institutional Custom GPTs): El siguiente paso tecnológico es superar el uso de chatbots generalistas (como ChatGPT o Gemini) para desarrollar agentes de IA institucionales basados en la arquitectura RAG (Retrieval-Augmented Generation).
    - Implementación: Entrenar o ajustar modelos ("fine-tuning") utilizando exclusivamente el corpus documental de la universidad (reglamentos, mallas curriculares, investigaciones previas y el PEDI).

- Propósito: Crear asistentes virtuales seguros que no "alucinen" información, sino que respondan a estudiantes y docentes basándose estrictamente en la normativa y el conocimiento institucional verificado, protegiendo la propiedad intelectual (Henriquez Orrego, 2025; Purificato et al., 2025).

B. Auditorías algorítmicas automatizadas: Para garantizar el cumplimiento continuo del Pilar I (Gobernanza), se debe avanzar hacia la automatización del control ético.
   - Implementación: Desplegar sistemas de monitoreo continuo que auditen los algoritmos utilizados en la gestión (ej. admisión, asignación de becas) para detectar desviaciones o sesgos discriminatorios en tiempo real.
   - Propósito: Pasar de la "ética declarativa" a la "ética verificable", generando reportes automáticos de cumplimiento para el Comité de Ética (CEIA) y las autoridades nacionales, consolidando la confianza pública en la gestión universitaria (OECD, 2025; PNUD, 2025b).
   - En definitiva, el Marco IA593 sienta las bases para que la universidad no sea arrastrada por la ola tecnológica, sino que la navegue con brújula ética, rigor científico y visión de estado.

## 10. Referencias Bibliográficas

Abbasnejad, B., Soltani, S., Taghizadeh, F., & Zare, A. (2025). Developing a multilevel framework for AI integration in technical and engineering higher education: insights from bibliometric analysis and ethnographic research. *Interactive Technology and Smart Education*. https://doi.org/10.1108/ITSE-12-2024-0314

Acuña Contreras, I., Cravero Leal, A., & Ferreira Vergara, A. (2024). Proposal for a Data Governance Framework in Higher Education. *2024 43rd International Conference of the Chilean Computer Science Society (SCCC)*, 1-9. https://doi.org/10.1109/SCCC63879.2024.10767672

Adeoye, O. O., Alimi, A. A., Agboola, O. S., Akindele, A. T., Arulogun, O. T., & Adigun, G. O. (2025). Advancing Higher Education through Artificial Intelligence (AI): A Framework for Teaching, Assessment, and Research Integration. *East African Journal of Education Studies*, *8*(2), 292-308. https://doi.org/10.37284/eajes.8.2.2946

An, Y., Yu, J. H., & James, S. (2025). Investigating the higher education institutions' guidelines and policies regarding the use of generative AI in teaching, learning, research, and administration. *International Journal of Educational Technology in Higher Education*, *22*(10), 1-23. https://doi.org/10.1186/s41239-025-00507-3

Azevedo, L., Robles, P., Best, E., & Mallinson, D. J. (2025). Institutional Policies on Artificial Intelligence in Higher Education: Frameworks and Best Practices for Faculty. *New Directions for Adult and Continuing Education*, *2025*(188), 70-78. https://doi.org/10.1002/ace.70013

Bellas, F., Ooge, J., Roddeck, L., Rashheed, H. A., Skenduli, M. P., Masdoum, F., Zainuddin, N. bin, Gori, J. N., Costello, E., Kralj, L., Dcosta, D. T., Katsamori, D., Neethling, D., Maat, S. ter, Saurabh, R., Alasgarova, R., Radaelli, E., Stamatescu, A., Blazic, A., … Obae, C. (2025). *Explainable AI in education: Fostering human oversight and shared responsibility*. https://op.europa.eu/publication/manifestation_identifier/PUB_EC0125086ENN

Bittle, K., & El-Gayar, O. (2025). Generative AI and Academic Integrity in Higher Education: A Systematic Review and Research Agenda. *Information*, *16*(4), 1-15. https://doi.org/10.3390/info16040296

Burneo-Arteaga, P., Lira, Y., Murzi, H., Balula, A., & Costa, A. P. (2025). Capability-based training framework for generative AI in higher education. *Frontiers in Education*, *10*, 1-17. https://doi.org/10.3389/feduc.2025.1594199

Cacho, R. M. (2024). Integrating Generative AI in University Teaching and Learning: A Model for Balanced Guidelines. *Online Learning Journal*, *28*(3), 55-81. https://doi.org/10.24059/olj.v28i3.4508

Castañeda de León, L. M., Ramírez Molina, A. Y., Castillejos Reyes, J. M., & Ventura Miranda, M. T. (2025). *Uso y desarrollo ético de la Inteligencia Artificial en la Universidad: docencia e investigación*. https://www.tic.unam.mx/wp-content/uploads/2025/11/Uso-y-desarrollo-etico-del-IA-en-la-UNAM_v-digital.pdf

Castelló-Sirvent, F., Roger-Monzó, V., & Gouveia-Rodrigues, R. (2024). Quo Vadis, University? A Roadmap for AI and Ethics in Higher Education. *Electronic Journal of e-Learning*, *22*(6), 34-51. https://doi.org/10.34190/ejel.22.6.3267

Cedeño Meza, J. G., Maitta Rosado, I. S., Vélez Zambrano, M. L., & Palomeque Zambrano, J. Y. (2024). University research with artificial intelligence. *Revista Venezolana de Gerencia*, *29*(106), 817-830. https://doi.org/10.52080/rvgluz.29.106.23

CENIA. (2024). *Índice Latinoamericano de Inteligencia Artificial (ILIA) 2024*. https://indicelatam.cl/

CENIA. (2025). *Índice Latinoamericano de Inteligencia Artificial (ILIA) 2025*. https://indicelatam.cl/

CES. (2023). *Reglamento de Régimen Académico*. https://gaceta.ces.gob.ec/resultados.html?id_documento=251023

Comité de Ética de Investigación en Seres Humanos de la Universidad Nacional de Loja. (2023). *Reglamento del Comité de Ética de Investigación en Seres Humanos de la Universidad Nacional de Loja*. https://unl.edu.ec/ceish

Cruz, A., Mora, R., Andonova, V., Rosales, C., Carrasco, C., & Castillo Leska, A. (2025). *fAIr Tech Radar. Explorando la adopción de inteligencia artificial en América Latina y el Caribe*. https://doi.org/10.18235/0013837

Díaz-Noguera, M. D., Hervás Gómez, C., Florina Grosu, E., Mâță, L., & Musata-Dacia, B. (Eds.). (2025). *Inteligencia artificial y educación: innovación pedagógica para un aprendizaje transformador* (1.ª ed.). Dykinson. https://www.dykinson.com/libros/inteligencia-artificial-y-educacion-innovacion-pedagogica-para-un-aprendizaje-transformador/9788410706958/

Elhajji, M., Alsayyari, A. S., & Alblawi, A. (2020). Towards an artificial intelligence strategy for higher education in Saudi Arabia. *2020 3rd International Conference on Computer Applications & Information Security (ICCAIS)*, 1-7. https://doi.org/10.1109/ICCAIS48893.2020.9096833

Fábrega Lacoa, R., Piña Pérez, K., Domínguez Caro, J., Fábrega Lacoa, J., Demattei, L., Manzur Schulz, S., & Brandt Garcés, J. (2024). *Oportunidades de Innovación Pedagógica con Asistencia de Inteligencia Artificial (IA)*. https://scioteca.caf.com/handle/123456789/2207

George, B., & Wooden, O. (2023). Managing the Strategic Transformation of Higher Education through Artificial Intelligence. *Administrative Sciences*, *13*(9), 1-20. https://doi.org/10.3390/admsci13090196

Giandana Gigena, F., Pisanu, G., Alarcón, Á., Sampieri, A., Mojica, Y., Leufer, D., & Escoto, W. (2024). *Radiografía Normativa: ¿dónde, qué y cómo se está regulando la Inteligencia Artificial en América Latina?* Access Now. https://raeia.org/books/informe-de-politicas-publicas-de-ia-en-america-latina/

Gobierno de Chile. (2021). *Política Nacional de Inteligencia Artificial*. https://www.minciencia.gob.cl/areas/inteligencia-artificial/politica-nacional-de-inteligencia-artificial/

González-Fernández, M. O., Romero-López, M. A., Sgreccia, N. F., & Latorre Medina, M. J. (2025). Normative framework for ethical and trustworthy AI in higher education: state of the art. *RIED-Revista Iberoamericana de Educacion a Distancia*, *28*(2), 181-208. https://doi.org/10.5944/ried.28.2.43511

Grupo de trabajo de Inteligencia Artificial Generativa de la UNAM. (2025). *Recomendaciones para el uso educativo de la Inteligencia Artificial Generativa en la UNAM*. https://www.ceide.unam.mx/index.php/recomendaciones-para-el-uso-educativo-de-la-inteligencia-artificial-generativa-en-la-unam/

Grupo sobre Inteligencia Artificial en la Educación Superior. (2025). *Hacia un marco institucional para apropiar la Inteligencia Artificial en Universidades Latinoamericanas: Lecciones del Grupo de Trabajo sobre la IA en la Educación Superior en América Latina*. https://thedialogue.org/analysis/hacia-un-marco-institucional-para-apropiar-la-inteligencia-artificial-en-universidades-latinoamericanas?lang=es

Guersenzvaig, A., & Monett, D. (2026). *Resisting Enchantment and Determinism: How to critically engage with AI university guidelines*. https://doi.org/10.5281/zenodo.18282338

Guizani, S., Mazhar, T., Shahzad, T., Ahmad, W., Bibi, A., & Hamam, H. (2025). A systematic literature review to implement large language model in higher education: issues and solutions. *Discover Education*, *4*, 1-25. https://doi.org/10.1007/s44217-025-00424-7

Guridi, J. A., & Trivelli, P. (2025). *Recomendaciones para la implementación de sandboxes regulatorios de IA*. https://publications.iadb.org/es/recomendaciones-para-la-implementacion-de-sandboxes-regulatorios-de-ia

Hasan, S., Amin, A., Tanim, S. H., Jin, Y., Zhu, G., & Ma, H. (2024). Perspectives on Artificial Intelligence Integration in Higher Education: Moral Implications and Data Privacy Concerns. *2024 10th International Conference on Computer and Communications (ICCC)*, (2024), 1516-1520. https://doi.org/10.1109/ICCC62609.2024.10942029

Henriquez Orrego, A. (2025). *Guía para crear GPTs personalizados*. https://raeia.org/books/guia-para-crear-gpts-personalizados/

Hinojosa Mamani, J., Catacora Lucana, E., & Mamani Gamarra, J. E. (2024). Bitácora de herramienta digitales: la inteligencia artificial en la investigación y las producciones académicas. En *BITÁCORA DE HERRAMIENTA DIGITALES: LA INTELIGENCIA ARTIFICIAL EN LA INVESTIGACIÓN Y LAS PRODUCCIONES ACADÉMICAS* (1.ª ed.). Editora Científica Digital. https://doi.org/10.37885/978-65-5360-555-8

IEEE Computer Society. (2026). *Technology Predictions 2026*. https://www.computer.org/resources/2026-tech-predictions

IESALC. (2021). *Pensar más allá de los límites: perspectivas sobre los futuros de la educación superior hasta 2050*. https://unesdoc.unesco.org/ark:/48223/pf0000377529

Iskandarova, S., Yusif-Zada, K., & Mukhtarova, S. (2024). Integrating AI Into Higher Education Curriculum in Developing Countries. *Proceedings - Frontiers in Education Conference, FIE*, 1-9. https://doi.org/10.1109/FIE61694.2024.10893097

Kakhkharova, M., & Tuychieva, S. (2024). AI-Enhanced Pedagogy in Higher Education: Redefining Teaching-Learning Paradigms. *2024 International Conference on Knowledge Engineering and Communication Systems, ICKECS 2024*, 1-6. https://doi.org/10.1109/ICKECS61492.2024.10616893

Le Dinh, T., Le, T. D., Uwizeyemungu, S., & Pelletier, C. (2025). Human-Centered Artificial Intelligence in Higher Education: A Framework for Systematic Literature Reviews. *Information*, *16*(3), 1-20. https://doi.org/10.3390/info16030240

Li, Y., Tolosa, L., Rivas-Echeverria, F., & Marquez, R. (2025). Integrating AI in Education: Navigating UNESCO Global Guidelines, Emerging Trends, and Its Intersection with Sustainable Development Goals. En *ChemRxiv* (ChemRxiv). https://doi.org/10.26434/chemrxiv-2025-wz4n9

Mahrishi, M., Abbas, A., & Siddiqui, M. K. (2024). Global Initiatives Towards Regulatory Frameworks for Artificial Intelligence (AI) in Higher Education. *Digital Government: Research and Practice*, *6*(2), 1-9. https://doi.org/10.1145/3672462

Masbernat, A.C., P., Cornejo- Plaza, I., & Cippitani, R. (2024). Scientific integrity in university education in the context of artificial intelligence. *Revista de Educación y Derecho*, (2-Extraordinario), 207-248. https://doi.org/10.1344/REYD2024.2-Extraordinario.49189

Maslej, N., Fattorini, L., Perrault, R., Gil, Y., Parli, V., Kariuki, N., Capstick, E., Reuel, A., Brynjolfsson, E., Etchemendy, J., Ligett, K., Lyons, T., Manyika, J., Carlos Niebles, J., Shoham, Y., Wald, R., Hamrah, A., Santarlasci, L., Betts Lotufo, J., … Oak, S. (2025). *The AI Index 2025 Annual Report*. https://doi.org/10.48550/arXiv.2504.07139

McLennan, S., Fiske, A., Celi, L. A., Müller, R., Harder, J., Ritt, K., Haddadin, S., & Buyx, A. (2020). An embedded ethics approach for AI development. *Nature Machine Intelligence*, *2*(9), 488-490. https://doi.org/10.1038/s42256-020-0214-1

Microsoft. (2025). *2025 AI in Education: A Microsoft Special Report*. https://www.microsoft.com/en-us/education/blog/2025/08/ai-in-education-report-insights-to-support-teaching-and-learning/

Ministry of National Education. (2025). *Artificial Intelligence in Education Policy Document and Action Plan (2025–2029)*. https://yegiteken.meb.gov.tr/www/the-artificial-intelligence-in-education-policy-document-and-action-plan-20252029-has-been-published-in-english/icerik/84

MINTEL. (2021). *Ley Orgánica de Protección de Datos Personales*. https://www.telecomunicaciones.gob.ec/wp-content/uploads/2023/11/LOPDP-LEXIS.pdf

MINTEL. (2023). *Reglamento General de la Ley Orgánica de Protección de Datos Personales*. https://www.telecomunicaciones.gob.ec/wp-content/uploads/2023/11/Decreto-Ejecutivo-No.-904.pdf

MINTEL. (2026). *Estrategia para el Fomento del Desarrollo y Uso Ético y Responsable de la Inteligencia Artificial en el Ecuador*. https://esacc.corteconstitucional.gob.ec/storage/api/v1/10_DWL_FL/eyJjYXJwZXRhIjoicm8iLCJ1dWlkIjoiMjRiYTA4MGYtNmY0Yy00OWVlLWFjMzAtYjg2ZDA4MTE3MTE1LnBkZiJ9?ref=youtopiaecuador.com

Molina, E., & Medina, E. (2025). *Revolución de la IA en Educación Superior. Lo que hay que saber*. https://www.bancomundial.org/es/region/lac/publication/ia-educacion-superior-inteligencia-artificial

MTDFP. (2025). *Guía 7. Datos y Gobernanza del dato*. https://marketplace.innovaciondespachos.com/community/blog/post/guias-oficiales-reglamento-europeo-inteligencia-artificial-espana

Naciones Unidas. (2025). *Informe sobre tecnología e innovación. Inteligencia artificial inclusiva para el desarrollo*. https://unctad.org/es/publication/informe-sobre-tecnologia-e-innovacion-2025

Naranjo Bautista, S., Alessandro, M., & Manuel Juan, O. D. Z. (2025). *Inteligencia artificial para Estados más efectivos*. https://doi.org/10.18235/0013786

OdiseIA. (2022). *GuIA de buenas prácticas en el uso de la inteligencia artificial ética*. https://www.pwc.es/es/publicaciones/tecnologia/odiseia-pwc-guia-responsable-ia.html

OECD. (2023a). *OECD Digital Education Outlook 2023: Towards an Effective Digital Education Ecosystem* (OECD Digital Education Outlook). OECD Publishing. https://doi.org/10.1787/c74f03de-en

OECD. (2023b). *The state of implementation of the OECD AI Principles four years on*. https://www.oecd.org/en/publications/the-state-of-implementation-of-the-oecd-ai-principles-four-years-on_835641c9-en.html

OECD. (2025). *Gobernar con la inteligencia artificial: Panorama actual y hoja de ruta en las funciones centrales de gobierno*. OECD Publishing. https://doi.org/10.1787/dc00e56a-es

OECD. (2026). *OECD Digital Education Outlook 2026: Exploring Effective Uses of Generative AI in Education* (OECD Digital Education Outlook). OECD Publishing. https://doi.org/10.1787/062a7394-en

OEI-UNESCO. (2025). *El Estado de la Ciencia. Principales Indicadores de Ciencia y Tecnología 2025*. https://oei.int/oficinas/argentina/publicaciones/el-estado-de-la-ciencia-principales-indicadores-de-ciencia-y-tecnologia-iberoamericanos-interamericanos-2025/

Pedreño Muñoz, A., González Gosálbez, R., Mora Illán, T., Pérez Fernández, E. del M., Ruiz Sierra, J., & Torres Penalva, A. (2024). *La inteligencia artificial en las universidades: retos y oportunidades* (1.ª ed.). Grupo 1MillionBot. https://1millionbot.com/la-inteligencia-artificial-en-las-universidades-retos-y-oportunidades/

Penabad-Camacho, L., Morera-Castro, M., & Penabad-Camacho, M. A. (2024). Guide for the use and reporting of Artificial Intelligence in scientific-academic journals. *Revista Electronica Educare*, *28*, 1-41. https://doi.org/10.15359/ree.28-S.19830

PNUD. (2025a). *AILA: Evaluación del Panorama de la Inteligencia Artificial en Ecuador*. https://www.undp.org/es/ecuador/publicaciones/evaluacion-del-panorama-de-inteligencia-artificial-ia

PNUD. (2025b). *Atlas de Inteligencia Artificial para el Desarrollo Humano de América Latina y El Caribe*. https://www.undp.org/es/latin-america/publicaciones/atlas-de-inteligencia-artificial-para-america-latina-y-el-caribe

Pradit, K., & Piriyasurawong, P. (2024). Architecture of Artificial Intelligence Entrepreneurship on Digital Ecosystem for Higher Education Institutions. *2024 9th International STEM Education Conference (iSTEM-Ed)*, 1-6. https://doi.org/10.1109/iSTEM-Ed62750.2024.10663150

Purificato, E., Bili, D., Jungnickel, R., Serra Ruiz, V., Fabiani, J., Abendroth Dias, K., Fernandez Llorca, D., & Gomez, E. (2025). *The Role of Artificial Intelligence in Scientific Research - A Science for Policy, European Perspective*. Publications Office of the European Union. https://doi.org/10.2760/3050242

Ramírez Téllez, A., Fonseca Ortiz, L. M., & Triana Domínguez, F. C. (2024). Artificial Intelligence in University Administration: An Overview of its Uses and Applications. *Revista Interamericana de Bibliotecologia*, *47*(2), 1-12. https://doi.org/10.17533/udea.rib.v47n2e353620

Rivas, A. (2025a). *La llegada de la IA a la educación en América Latina: en construcción*. https://oei.int/oficinas/secretaria-general/publicaciones/la-llegada-de-la-ia-a-la-educacion-en-america-latina-en-construccion/

Rivas, A. (2025b). *La llegada de la IA a la educación superior en Iberoamérica: Un mapa para diseñar estrategias institucionales*. https://oei.int/oficinas/secretaria-general/publicaciones/la-llegada-de-la-ia-a-la-educacion-superior-en-iberoamerica/

Rivas, A., Buchbinder, N., & Barrenechea, I. (2023). *El futuro de la Inteligencia Artificial en educación en América Latina*. https://profuturo.education/observatorio/tendencias/el-futuro-de-la-inteligencia-artificial-en-educacion-en-america-latina/

Rivas, P., Ortiz, J., Díaz-Pachón, D. A., & Montoya, L. N. (2023). Bridging Industry, Government, and Academia for Socially Responsible AI: The CSEAI Initiative. *2023 IEEE International Symposium on Ethics in Engineering, Science, and Technology (ETHICS)*, 1-1. https://doi.org/10.1109/ETHICS57328.2023.10155071

Ruiz Muñoz, G. F., Vasco Delgado, J. C., & Alvear Dávalos, J. M. (2024). Artificial intelligence and governance in the academic and administrative management of higher education. *Revista Social Fronteriza*, *4*(6), 1-19. https://doi.org/10.59814/resofro.2024.4(6)e508

Salas, J., Patterson, G., & De Barros Vidal, F. (2022). A Systematic Mapping of Artificial Intelligence Solutions for Sustainability Challenges in Latin America and the Caribbean. *IEEE Latin America Transactions*, *20*(11), 2312-2329. https://doi.org/10.1109/TLA.2022.9904756

Serrano Acitores, A. (2025). *La Inteligencia Artificial Generativa en el Trabajo de Fin de Grado: principios para una integración transparente y responsable en la Universidad Española*. https://papers.ssrn.com/sol3/papers.cfm?abstract_id=5933434

Sieker, F., Tarkowski, A., Gimpel, L., & Osborne, C. (2025). *Public AI White Paper – A Public Alternative to Private AI Dominance*. https://doi.org/10.11586/2025040

Southworth, J., Migliaccio, K., Glover, J., Glover, J. N., Reed, D., McCarty, C., Brendemuhl, J., & Thomas, A. (2023). Developing a model for AI Across the curriculum: Transforming the higher education landscape via innovation in AI literacy. *Computers and Education: Artificial Intelligence*, *4*, 1-10. https://doi.org/10.1016/j.caeai.2023.100127

Temper, M., Tjoa, S., & David, L. (2025). Higher Education Act for AI (HEAT-AI): a framework to regulate the usage of AI in higher education institutions. En *Frontiers in Education* (Vol. 10). Frontiers Media SA. https://doi.org/10.3389/feduc.2025.1505370

The Institute of Internal Auditors. (2023). *Marco de Auditoría de Inteligencia Artificial*. https://www.theiia.org/en/content/tools/professional/2023/the-iias-updated-ai-auditing-framework/

UNESCO. (2023a). *ChatGPT e Inteligencia Artificial en la educación superior. Guía de inicio rápido*. https://unesdoc.unesco.org/ark:/48223/pf0000385146_spa

UNESCO. (2023b). *Oportunidades y desafíos de la era de la inteligencia artificial para la educación superior. Una introducción para los actores de la educación superior*. https://unesdoc.unesco.org/ark:/48223/pf0000386670_spa

UNESCO. (2024). *Guía para el uso de IA generativa en educación e investigación*. https://unesdoc.unesco.org/ark:/48223/pf0000389227_spa

UNESCO. (2025a). *Ecuador. Evaluación del estadio de preparación en materia de Inteligencia Artificial (IA) de la Unesco*. https://unesdoc.unesco.org/ark:/48223/pf0000396465_spa

UNESCO. (2025b). *Los desafíos de la IA en la educación superior y las respuestas institucionales: ¿hay espacio para marcos de competencias?* https://unesdoc.unesco.org/ark:/48223/pf0000394935_spa.locale=en

UNESCO. (2025c). Marco de competencias para docentes en materia de IA. En *Marco de competencias para docentes en materia de IA*. UNESCO. https://doi.org/10.54675/AQKZ9414

UNESCO. (2025d). Marco de competencias para estudiantes en materia de IA. En *Marco de competencias en materia de IA para estudiantes*. UNESCO. https://doi.org/10.54675/EKCU4552

UNESCO. (2025e). *PERÚ: Evaluación del estadio de preparación de la inteligencia artiicial*. https://unesdoc.unesco.org/ark:/48223/pf0000393824_spa

Universidad Andrés Bello. (2024). *Lineamiento para el uso responsable de inteligencia artificial en la Universidad Andrés Bello*. https://ia.unab.cl/lineamientos/

Universidad Europea. (2024). *Observatorio de Inteligencia Artificial en Educación Superior*. https://universidadeuropea.com/conocenos/observatorio-inteligencia-artificial-educacion-superior/

Universidad Nacional de Loja. (2025). *Plan Estratégico de Desarrollo Institucional UNL 2024-2028. UNL Sostenible*. https://unl.edu.ec/pedi/2024-2028

Universidad Pontificia Comillas. (2024). Guía práctica de aplicación de la IA. En *Guía práctica de aplicación de la IA*. Universidad Pontificia Comillas. https://doi.org/10.14422/OAID20241126

Villegas Dianta, A., Henríquez Mardónez, A., Henríquez Orrego, A., Correa Concha, B., Bustamante Olivares, B., Sepúlveda Irribarra, C., Sandoval Sepúlveda, C., Pica Miranda, G., Díaz Fernández, J. I., González Ayala, M., Rondon Cabrera, O., Espejo Aubá, P., Figueroa Ayala, P., & Monge Rogel, R. (2025). *Marco para el uso de la Inteligencia Artificial en UDLA. Docencia, Investigación y Vinculación con el Medio*. https://www.udla.cl/descargas/normativas/marco-para-uso-ia-en-udla.pdf

Vivas Urias, M. D., & Ruiz Rosillo, M. A. (2025). *Inteligencia artificial generativa. Buenas prácticas docentes en educación superior* (1.ª ed.). OCTAEDRO, S.L. https://doi.org/10.36006/09648-1

Wang, J., Huo, Y., Mahe, J., Ge, Z., Liu, Z., Wang, W., & Zhang, L. (2024). Developing an Ethical Regulatory Framework for Artificial Intelligence: Integrating Systematic Review, Thematic Analysis, and Multidisciplinary Theories. *IEEE Access*, *12*, 179383-179395. https://doi.org/10.1109/ACCESS.2024.3501332

WEF. (2026). *Latin America in the Intelligent Age: A New Path for Growth*. https://www.weforum.org/publications/latin-america-in-the-intelligent-age-a-new-path-for-growth/

WOOCLAP. (2025). *IA para la Educación Superior. 10 universidades comparten sus mejores prácticas*. https://www.getwooclap.com/es/whitepaper-ia

Yu, X., Xia, J., & Shen, Z. (2024). Artificial Intelligence for Research (AI4R): Knowledge Base and Impact. *2024 International Conference on Intelligent Education and Intelligent Research (IEIR)*, 1-8. https://doi.org/10.1109/IEIR62538.2024.10959986

## 11. Declaración de delegación a Inteligencia Artificial Generativa

Los autores declaran el uso de inteligencia artificial generativa en el proceso de investigación y redacción. De acuerdo con la taxonomía GAIDeT (2025), las siguientes tareas fueron delegadas a herramientas de IAG bajo plena supervisión humana: redacción de la revisión de la literatura, evaluación de la novedad de la investigación e identificación de brechas, recolección de datos, curaduría y organización de datos, generación de texto, corrección y edición de estilo, síntesis de textos, formulación de conclusiones, traducción, reformateo, análisis de riesgos éticos, identificación de tendencias, identificación de limitaciones.

La herramienta de IAG utilizada fue: NotebookLM.

La responsabilidad del manuscrito final recae íntegramente en los autores.

Las herramientas de IAG no se incluyen como autores ni asumen responsabilidad sobre los resultados finales.

Declaración presentada por: Los investigadores.